\documentclass[11pt,letterpaper]{article}
\pdfoutput=1

\usepackage{jcappub}
\usepackage{amsmath}
\usepackage{amsfonts}
\usepackage{amssymb}
\usepackage{graphicx}
\usepackage{epsfig}
\usepackage{color}
\usepackage{multirow}
\usepackage{graphicx}
\usepackage{bigints}
\usepackage{todonotes,bm,toolbox}
\usepackage{calligra}
\usepackage{hyperref}
\usepackage[normalem]{ulem}
\usepackage{afterpage}

\def\ba{\begin{eqnarray}}
\def\ea{\end{eqnarray}}
\def\bea{\begin{eqnarray}}
\def\eea{\end{eqnarray}}
\def\be{\begin{equation}}
\def\ee{\end{equation}}
\def\d{\mathrm{d}}
\def\nn{\nonumber}
\def\({\left(}
\def\){\right)}
\def\[{\left[}
\def\]{\right]}
\DeclareMathOperator{\tr}{tr}

\def\eps{\epsilon}

\def\comment#1{}
\renewcommand{\v}[1]{\bm{#1}}

\newcommand{\cs}{\,,\  } 

\def\vx{\v{x}}
\def\vk{\v{k}}

\def\vp{\v{p}}

\def\lapl{{\nabla^2}}
\def\bI{\beta_{\lapl\v{v}}}
\def\bII{\beta_{\partial_\parallel^2 \v{v}}}

\def\epsv{\varepsilon}

\newcommand{\refeq}[1]{eq.~(\ref{eq:#1})}

\title{Information content in the redshift-space galaxy power spectrum and bispectrum}

\author[a]{Nishant Agarwal,}
\author[b]{Vincent Desjacques,}
\author[c]{Donghui Jeong,}
\author[d]{and Fabian Schmidt}

\affiliation[a]{Department of Physics and Applied Physics, University of Massachusetts, Lowell, MA 01854, USA}
\affiliation[b]{Physics department, Technion, 3200003 Haifa, Israel}
\affiliation[c]{Department of Astronomy and Astrophysics and Institute for Gravitation and the Cosmos, The Pennsylvania State University, University Park, PA 16802, USA}
\affiliation[d]{Max-Planck-Institut f\"ur Astrophysik, Karl-Schwarzschild-Stra\ss e~1, 85748 Garching, Germany}

\emailAdd{nishant\_agarwal@uml.edu}
\emailAdd{dvince@physics.technion.ac.il}
\emailAdd{djeong@psu.edu}
\emailAdd{fabians@mpa-garching.mpg.de}

\date{\today}

\abstract{We present a Fisher information study of the statistical impact of galaxy bias and selection effects on the estimation of key cosmological parameters from galaxy redshift surveys; in particular, the angular diameter distance, Hubble parameter, and linear growth rate at a given redshift, the cold dark matter density, and the tilt and running of the primordial power spectrum. The line-of-sight-dependent selection contributions we include here are known to exist in real galaxy samples. We determine the maximum wavenumber included in the analysis by requiring that the next-order corrections to the galaxy power spectrum or bispectrum, treated here at next-to-leading and leading order, respectively, produce shifts of $\lesssim 0.25\sigma$ on each of the six cosmological parameters. With the galaxy power spectrum alone, selection effects can deteriorate the constraints severely, especially on the linear growth rate. Adding the galaxy bispectrum helps break parameter degeneracies significantly. We find that a joint power spectrum-bispectrum analysis of a Euclid-like survey can still measure the linear growth rate to 10\% precision after complete marginalization over selection bias. We also discuss systematic parameter shifts arising from ignoring selection effects and/or other bias parameters, and emphasize that it is necessary to either control selection effects at the percent level or marginalize over them. We obtain similar results for the Roman Space Telescope and HETDEX.}

\begin{document}
\maketitle


\section{Introduction}

Cosmological perturbation theory has been extremely successful in explaining cosmic structures on large scales (that is, in the linear regime), including temperature anisotropies and polarization of the cosmic microwave background (CMB) and statistics of the large-scale distribution of galaxies (large-scale structure, LSS). Combined with rich observational datasets \cite{Bennett:2012zja,Alam:2016hwk,Abbott:2017wau,Aghanim:2018eyx}, it allows us to measure most cosmological parameters in the concordance $\Lambda$-cold dark matter ($\Lambda$CDM) model to better than $1\%$ accuracy. In the case of LSS, going beyond the linear regime is expected to provide significant improvements in parameter constraints, specifically on the expansion history, growth rate of cosmic structures, statistical properties of the initial fluctuations, and mass of neutrinos.

In practice, besides the weak gravitational lensing tomography, the three-dimensional distribution of LSS can only be observed indirectly from observations of {\it biased} LSS tracers, such as galaxies, clusters of galaxies, or the intensity mapping of emission lines, for example Lyman-$\alpha$ or $21~{\rm cm}$. For a review of nonlinear perturbation theory techniques in LSS see \cite{Bernardeau:2001qr} and for a review of galaxy bias see \cite{Desjacques:2016bnm}. Furthermore, observations are made in redshift space (that is, inferring the distances by observed spectral shift) and in general involve line-of-sight dependent selection effects. The latter are particularly important for observables based on resonance lines, such as the Lyman-$\alpha$ forest, where the probability of detecting an emitted photon depends, in particular, on the velocity gradient along the line of sight \cite{Zheng:2010jf,Wyithe:2011mt,Behrens:2017xmm}. The selection effect, however, can also be relevant for galaxies selected on photometric properties. For example, galaxies tend to align with large-scale tidal fields, and the flux and thus selection of galaxies in general depends on their orientation with respect to the line of sight \cite{Hirata:2009qz,Krause:2010tt,Martens:2018uqj,Obuljen:2019ukz}. Such selection effects are not included in the standard bias expansion \cite{Desjacques:2016bnm}.

Building on significant work in the literature over the past 20 years, a complete description of the observed galaxy statistics, including all physical effects mentioned above, has finally been assembled recently. In \cite{Desjacques:2018pfv} a few of us found, using an effective field theory (EFT) approach (see \cite{Porto:2016pyg} for a review) to generate all possible perturbative contributions \cite{Baumann:2010tm,Carrasco:2012cv,Senatore:2014eva,Senatore:2014vja,Mirbabayi:2014zca}, that selection effects come in as various counterterms required to consistently renormalize observables such as the redshift-space power spectrum and bispectrum. As a result, a complete description of the galaxy power spectrum at 1-loop order and the galaxy bispectrum at tree-level requires a total of 22 parameters, that include (i) 5 galaxy bias parameters and 5 rest-frame stochastic amplitudes, (ii) 9 selection parameters, and (iii) 3 velocity bias parameters, one of which is due to selection and one is a stochastic parameter.

Selection effects can be degenerate with cosmological parameters. For example, the parameter $b_{\eta}$ (defined in section\ \ref{sec:reviewmatter}) is perfectly degenerate with the linear Kaiser effect \cite{Kaiser:1987qv}. As a consequence, the leading-order galaxy power spectrum alone cannot be used to constrain the linear growth rate $f(z)$. The 1-loop contribution to the galaxy power spectrum and the tree-level galaxy bispectrum, on the other hand, show much richer wavevector dependences which may break the degeneracies amongst various bias parameters. At the same time, these higher-order contributions come with the price of including the aforementioned 22 bias parameters that need to be marginalized over. The goal of this paper is to quantify the extent to which selection effects degrade cosmological constraints, how the galaxy bispectrum helps in mitigating degeneracies, and how parameter constraints shift in the presence of model systematics. We focus on measurements of the following six cosmological parameters: the angular diameter distance $D_A(z)$ to galaxies at redshift $z$, Hubble expansion rate $H(z)$, current CDM density $\Omega_{c0}$, linear growth rate $f(z)$, and tilt $n_s$ and running $n_{\rm run}$ of the primordial power spectrum.

We construct a Fisher information matrix in the 22 parameters that describe the galaxy distribution and the six cosmological parameters. We determine the largest wavenumbers $k_{\rm max}$ that can be used in the galaxy power spectrum and bispectrum measurements from the requirement that the next-order perturbative contributions neglected in our model (that is, the 2-loop galaxy power spectrum and 1-loop galaxy bispectrum) do not systematically bias the best-fit values of any of the six cosmological parameters by more than $0.25\sigma$.

\subsection*{Summary of results}

Ignoring selection effects we find that, for a Euclid-like survey \cite{Euclid} with mean redshift $z=1.4$, the linear growth rate is weakly degenerate with other model parameters and can be measured at the few-percent level from the 1-loop galaxy power spectrum alone. Adding the tree-level galaxy bispectrum reduces the uncertainty by about a factor of 4. On including line-of-sight selection effects, however, we loose all constraining power from the galaxy power spectrum alone (the marginalized error on $f$ is of the order of $f$ itself). Interestingly, a joint power spectrum-bispectrum analysis helps alleviate many of the degeneracies with selection effects and allows us to measure $f$ at the $10\%$ level. We also quantify the extent to which model parameters shift if the true value of selection biases differ from their fiducial value. For $f$ in particular, we find that, if the lowest-order selection bias $b_{\eta}$ is fixed to a value that differs from the true value by only $4\%$, the growth rate inferred from the power spectrum alone is systematically biased by $\gtrsim 1\sigma$. Upon including the bispectrum, even a $\sim 1\%$ systematic error in $b_\eta$ can induce a $1\sigma$ shift if the parameter is fixed. We obtain similar results for the Roman Space Telescope\footnote{Formerly known as the Wide-Field InfraRed Space Telescope (WFIRST).} \cite{Green:2012mj} and HETDEX \cite{Hill:2008mv}. \\

The paper is organized as follows. We briefly review nonlinear matter clustering, galaxy bias, and selection effects in section\ \ref{sec:reviewmatter} and summarize the expressions for the 1-loop galaxy power spectrum and tree-level galaxy bispectrum including selection effects in section \ref{sec:reviewgalaxy}. In section\ \ref{sec:fisher} we set up the Fisher matrix and parameter shifts calculations. In section\ \ref{sec:fisherresults} we present the results of our Fisher analysis for Euclid, the Roman Space Telescope, and HETDEX. We conclude in section\ \ref{sec:conc}. In the first two appendices we summarize the calculation of Fisher matrix elements for the six cosmological parameters considered here and in the third appendix we detail the Fisher results for Euclid. 


\section{Large-scale galaxy distribution}
\label{sec:reviewmatter}

This section provides a succinct overview of the perturbative approach to describing the nonlinear distribution of the matter and galaxy density fields. We will ignore baryon-CDM relative density and velocity perturbations, combining both components into a single matter component, and assume Gaussian initial conditions (ignoring any primordial non-Gaussianity). We will further include the effect of massive neutrinos only through the linear matter power spectrum. We denote the matter overdensity as $\delta(\vx,\tau) = \rho(\vx,\tau)/\bar{\rho}(\tau) - 1$, where $(\vx,\tau)$ are conformal coordinates and $\rho$ is the total matter density, with $\bar{\rho}$ its homogeneous part. The nonlinear equations that govern the evolution of matter can be obtained by taking moments of the collisionless Boltzmann equation \cite{Bernardeau:2001qr,V2}; the zeroth- and first-order velocity moments give the continuity and Euler equations,
\bea
	\frac{\partial \delta(\vx,\tau)}{\partial \tau} + \partial_i \[ \{ 1 + \delta(\vx,\tau) \} v^i(\vx,\tau) \] & = & 0 \, ,
\label{eq:continuity} \\
	\frac{\partial v^i(\vx,\tau)}{\partial \tau} + {\cal H}(\tau) v^i(\vx,\tau) + v^j(\vx,\tau) \partial_j v^i(\vx,\tau) & = & -\partial^i \phi(\vx,\tau) \, ,
\label{eq:euler}
\eea
where $\v{v}(\vx,\tau) = d\vx/d\tau$ is the peculiar velocity of the fluid, $\phi$ is the gravitational potential, ${\cal H}\equiv aH$ is the conformal Hubble parameter, and the indices $i$, $j$ run over spatial components. Repeated indices imply summation. The Poisson equation relates the gravitational potential and matter fluctuations. This allows us to express the tidal shear as
\bea
	\Pi_{ij}^{[1]}(\vx,\tau) & \equiv & \frac{2}{3\Omega_m(\tau){\cal H}^2(\tau)} \partial_i \partial_j \phi(\vx,\tau) ,
\label{eq:Piij1}
\eea
where $\Omega_m$ is the matter fraction, and the density as $\delta=\tr\Pi^{[1]}$. In the Euler equation above, we have neglected the gradient of the stress tensor on the right-hand side. In an EFT approach, this term would give rise to a sound speed for the fluctuations, a bulk and shear viscosity, and a stochastic pressure component \cite{Baumann:2010tm,Carrasco:2012cv}. In this paper we will exclusively consider galaxy statistics. In this case, the EFT parameters for matter can be absorbed into galaxy bias coefficients. Note that this would change if one were to include the galaxy-matter cross power spectrum \cite{Desjacques:2016bnm}.

We can solve eqs.\ (\ref{eq:continuity}) and (\ref{eq:euler}) perturbatively. Transforming to Fourier space\footnote{We will use the Fourier convention $f(\vx) \, = \, \int_{\vk} e^{i\vk \cdot \vx} f \big( \vk \big)$, with the shorthand $\int_{\vk} \equiv \int \frac{d^3 k}{(2\pi)^3}$, throughout this paper.} and expanding $\delta$ in powers of the linear density fluctuation $\delta_L$ gives
\bea
	\delta (\vk,\tau) & = & \sum_{n = 1}^{\infty} D^n(\tau) \int_{\v{p}_1} \ldots \int_{\v{p}_n} \, (2\pi)^3 \delta_D ( \vk - \v{p}_{1 \ldots n} ) \delta_L(\v{p}_1) \ldots \delta_L(\v{p}_n) F_n(\v{p}_1, \ldots , \v{p}_n) \, , \quad
\label{eq:delta3}
\eea
where $D(\tau)$ is the linear growth factor, $\delta_D$ denotes a Dirac $\delta$-function, $\v{p}_{1 \ldots n} = \v{p}_1 + \ldots + \v{p}_n$, and the functions $F_n$ are the symmetrized standard perturbation theory (SPT) kernels, expressions for which can be found in, for example, \cite{Goroff:1986ep,Heavens:1998es}. Note that the mean matter overdensity in real space, calculated from eq.\ (\ref{eq:delta3}), vanishes: $\langle \delta (\vx,\tau) \rangle = 0$. We have also adhered to the commonly used assumption here that the kernels are time-independent and can be calculated in an Einstein-de Sitter Universe; the time-dependence comes in through powers of $D(\tau)$, where $D(\tau)$ is calculated for the actual cosmology \cite{Bernardeau:2001qr,Takahashi:2008yk}. This is quite accurate for standard $\Lambda$CDM and quintessence cosmologies \cite{Takahashi:2008yk}. Further, since we assume an irrotational fluid,\footnote{The vorticity induced by nonlinear structure is effectively a third-order term, which enters the bias expansion via the antisymmetric part $\partial_{\{i} v_{j\}}$ of the velocity shear. It also appears in the redshift-space galaxy density by contributing to $\partial_\parallel v_\parallel$. However, since the curl component of the velocity does not correlate with the density due to parity invariance, the former only contributes through auto-correlations of cubic operators and hence is of 2-loop or higher order \cite{Carrasco:2013mua,Mirbabayi:2014zca}.} the velocity divergence $\theta = \bm{\nabla}\cdot\bm{v}$ completely specifies the velocity field. 
The former is similarly expanded in powers of $\delta_L$,
\bea
	\theta (\vk,\tau) & = & -f(\tau) {\cal H}(\tau) \sum_{n = 1}^{\infty} D^n(\tau) \int_{\v{p}_1} \ldots \int_{\v{p}_n} \, (2\pi)^3 \delta_D \( \vk - \v{p}_{1 \ldots n} \) \delta_L(\v{p}_1) \ldots \delta_L(\v{p}_n) \nn \\
	& & \qquad \times \, G_n(\v{p}_1, \ldots , \v{p}_n) \, ,
\label{eq:theta3}
\eea
where $f(\tau) \equiv d \ln D/d \ln a$ and expressions for the functions $G_n$ can also be found in, for example, \cite{Heavens:1998es}.

Using eq.\ (\ref{eq:delta3}) we can now obtain the matter power spectrum and bispectrum, defined through
\bea
	\langle \delta(\vk,\tau) \delta(\v{p},\tau) \rangle & = & (2\pi)^3 \delta_D (\vk+\v{p}) P_{m}(k,\tau) \, , \\
	\langle \delta(\vk_1,\tau) \delta(\vk_2,\tau) \delta(\vk_3,\tau) \rangle & = & (2\pi)^3 \delta_D (\vk_1+\vk_2+\vk_3) B_{m}(\vk_1,\vk_2,\vk_3,\tau) \, .
\eea
At 1-loop order these are given by
\bea
	P_m(k,\tau) & = & P_L(k,\tau) + P_m^{2-2}(k,\tau) + 2P_m^{1-3}(k,\tau) \, , \\
	B_m(\vk_1,\vk_2,\vk_3,\tau) & = & B_m^{\rm LO}(\vk_1,\vk_2,\vk_3,\tau) + B_m^{\rm NLO}(\vk_1,\vk_2,\vk_3,\tau) \, ,
\eea
with $\langle \delta_L(\vk) \delta_L(\v{p}) \rangle = (2\pi)^3 \delta_D (\vk+\v{p}) P_L(k,\tau)/D^2(\tau)$,
\bea
	P_m^{2-2}(k,\tau) & = & 2 \int_{\v{p}} P_L(p,\tau) P_L(|\vk-\v{p}|,\tau) \[ F_2 (\v{p},\vk-\v{p}) \]^2 , \\
	P_m^{1-3}(k,\tau) & = & 3 P_L(k,\tau) \int_{\v{p}} P_L(p,\tau) F_3 (\v{p},-\v{p},\vk) \, ,
\eea
and
\bea
	B_m^{\rm LO}(\vk_1,\vk_2,\vk_3,\tau) & = & 2 P_L(k_1,\tau) P_L(k_2,\tau) F_2 (\vk_1,\vk_2) + {\rm 2 \ perm.} \, , \\
	B_m^{\rm NLO}(\vk_1,\vk_2,\vk_3,\tau) & = & B_m^{222} + B_m^{321,I} + B_m^{321,II} + B_m^{411} \, ,
\eea
where LO and NLO stand for leading-order and next-to-leading-order, the permutations refer to cyclic permutations in $(\vk_1,\vk_2,\vk_3)$, and we refer the reader to \cite{Bernardeau:2001qr} for explicit expressions of components of the 1-loop bispectrum. In the following sections we will suppress the time-dependence of quantities introduced in this section.

We now turn to the galaxy density field, beginning with the density in the galaxy rest-frame, that is, ignoring redshift-space distortions (RSDs) or selection effects. Following the EFT approach, the galaxy density is expanded via a general bias expansion
\ba
	\delta_g(\vx,\tau) & = & \sum_O \left[b_O(\tau) + \eps_O(\vx,\tau) \right][O](\vx,\tau) + \eps(\vx,\tau) \, ,
\label{eq:bias1}
\ea
where the sum runs over a list of operators $O$ (statistical fields) that are successively higher order in perturbations (and spatial derivatives). Ref.~\cite{Mirbabayi:2014zca} provides a convenient way to construct the complete bias expansion in terms of the density and tidal field, and their convective time derivatives, which together comprise the complete set of local gravitational observables, which we shall use in the following. As we have defined the equations with the renormalized operators $[{O}]$, all bias coefficients in \refeq{bias1} are observables, instead of the coefficients in the bare bias expansion \cite{Desjacques:2016bnm}. The coefficients $b_O$ are the deterministic bias parameters. The fields $\eps$ and $\eps_O$ are stochastic amplitudes which do not correlate with $\delta_m$ or any of the operators $O$. Their statistics, which asymptote to constants (white noise) as $k \to 0$ on large scales, are to be determined from the data, in a similar way as the $b_O$ \citep{Paech:2016hod}. Ref.~\citep{Hamaus:2010im,Ginzburg:2017mgf} have extracted the stochastic bias parameters from numerical simulations.

Each of the operators in \refeq{bias1} has a well-defined associated kernel $F_{O,n}$ at $n^{\rm th}$ order in perturbations. This allows us to write the galaxy density in the rest-frame in analogy to \refeq{delta3} as
\bea
	\delta_g (\vk,\tau) & = & \sum_{n = 1}^{\infty} D^n(\tau) \int_{\vp_1} \ldots \int_{\vp_n} \, (2\pi)^3 \delta_D ( \vk - \v{p}_{1 \ldots n} ) \delta_L(\v{p}_1) \ldots \delta_L(\v{p}_n) F_n^{(g)}(\v{p}_1, \ldots , \v{p}_n) \nonumber\\
&& + \mbox{ stochastic terms}\, , \qquad \ \ 
\label{eq:deltag}
\eea
where the galaxy kernels $F_n^{(g)}$ now also depend on various bias parameters.

So far, we have included all local gravitational observables from the point of view of an observer in a given galaxy. These do not make reference to the line of sight $\hat{\v{n}}$ that connects us to the galaxy. However, the number of galaxies we observe can depend on additional quantities depending on the details of how they are selected. For example, for galaxy surveys with a fixed depth, the galaxy selection function depends on all physical conditions affecting the observed line flux from each galaxy. In particular, the photon escape fraction is determined by the optical depth along the line of sight, which is a strong function of, among others, the column density of the absorber and the velocity gradient along the line of sight. In this case, one should expect an additional dependence on the velocity gradient projected along the line of sight \cite{Zheng:2010jf,Wyithe:2011mt,Behrens:2017xmm},
\bea
	\eta & \equiv & \partial_\parallel u_\parallel \, ,
\eea
where $\v{u} = \v{v}/{\cal H}$ and the subscript $\parallel$ stands for the line-of-sight component; for example, $\partial_\parallel \equiv \hat{n}^i \partial_i$. A large-scale tidal field may also impact the galaxy detection probability and lead to a preferred orientation relative to the line-of-sight \cite{Hirata:2009qz,Krause:2010tt,Martens:2018uqj,Obuljen:2019ukz}, leading to a dependence on
\bea
	\Pi^{[1]}_\parallel & \equiv & \Pi^{[1]}_{ij} \hat{n}^i \hat{n}^j\,,
\eea
which is simply proportional to $\eta$ at linear order in perturbations.

Such effects can be taken into account fully generally in the EFT approach by allowing for the line-of-sight to appear as a preferred direction in the bias expansion. The complete set of corresponding terms up to third order was presented in \cite{Desjacques:2018pfv},
\bea
\mbox{selection:} \quad & {\rm 1^{st}} \ & \ \Pi^{[1]}_\parallel \, , \nn
    \label{eq:BasisSel} \\ [3pt]
    & {\rm 2^{nd}} \ & \ \tr(\Pi^{[1]}) \Pi^{[1]}_\parallel \, , \
    (\Pi^{[1]} \Pi^{[1]})_\parallel 
    \cs
    \left(\Pi^{[1]}_\parallel \right)^2 \cs
    \Pi^{[2]}_\parallel \, , \nn \\ [3pt]
    & {\rm 3^{rd}} \ &
    \  \Pi_\parallel^{[1]}\tr(\Pi^{[1]} \Pi^{[1]}) \, ,
    \Pi_\parallel^{[1]}(\tr(\Pi^{[1]}))^2\,,
    (\Pi^{[1]} \Pi^{[1]})_\parallel\tr(\Pi^{[1]}) \, ,
    (\Pi^{[1]}\Pi^{[1]}\Pi^{[1]})_\parallel \, ,
    \nn \\ [3pt]
    & & \ 
    \left(\Pi_\parallel^{[1]}\right)^2\tr(\Pi^{[1]}) \, ,
    \Pi_\parallel^{[1]}(\Pi^{[1]}\Pi^{[1]})_\parallel \, ,
    \left(\Pi^{[1]}_\parallel\right)^3 ,
    \nn \\ [3pt]
    & & \     
    \tr(\Pi^{[1]}) \Pi^{[2]}_\parallel\cs
    (\Pi^{[1]} \Pi^{[2]})_\parallel \cs
    \Pi^{[2]}_\parallel  \Pi^{[1]}_\parallel\,,\
    \Pi^{[3]}_\parallel
    \,.
\eea
For $n>1$, the tensor $\Pi_{ij}^{[n]}$ is iteratively constructed from $\Pi_{ij}^{[n-1]}$ by applying convective time derivatives, and starts at $n^{\rm th}$  order in perturbation theory \cite{Mirbabayi:2014zca,Desjacques:2016bnm}. The subscript $\parallel$ stands for the line-of-sight component of the tensor as before; for example, $(A B)_\parallel \equiv A_{ij} B^j{}_k \hat{n}^i \hat{n}^k$.

Each selection term in \refeq{BasisSel} comes with a bias parameter, and the main goal of this paper is to explore the quantitative impact of these selection effects on the cosmology inference from galaxy redshift surveys. The first-order selection term is the most important one, as we will see that it is completely degenerate with the linear-order contribution from RSDs. It can be equivalently replaced by $\eta$, since $\eta$ and $\Pi^{[1]}_\parallel$ are directly proportional at linear order, while the differences at higher order in perturbations are absorbed by the remaining bias and selection terms.

In order to transform the galaxy distribution to redshift space we further need an expression for the galaxy velocity. As argued in \cite{Senatore:2014eva,Mirbabayi:2014zca} (also see \cite{Desjacques:2009kt}), the galaxy velocity cannot differ from the matter velocity by a multiplicative velocity bias. This is due to the equivalence principle: the relative velocity between galaxies and matter is a local observable and thus can be expanded in terms of the same fields as appear in the perturbative galaxy bias expansion itself, such as the matter density field. Then in order to obtain a vector quantity we need to take a spatial derivative, which shows that velocity bias involves at least two additional spatial derivatives on the velocity.

The expansion of the galaxy velocity including selection effects was also derived in \cite{Desjacques:2018pfv}. Consistently with our higher-derivative expansion for bias and selection contributions, we will keep only the leading contribution to velocity bias, which can be written as
\bea
	\v{v}_g & = & \v{v} + \bI \lapl\v{v} + \bII \partial_\parallel^2 \v{v} + \v{\epsv}_v(\vx,\tau) \, .
\label{eq:vg}
\eea
Selection effects lead to the third term in \refeq{vg}. Following the above arguments, the stochastic field $\v{\epsv}_v(\vk)$ in Fourier space is proportional to $k$ in the low-$k$ limit (we denote stochastic fields appearing in the galaxy velocity as $\epsv$, while those appearing in the density are denoted as $\eps$). 

Finally, having described the galaxy bias expansion in the galaxy rest-frame including selection effects, along with the bias relation for the galaxy velocity field, we can map the observed galaxy density into redshift space. The coordinate transformation is given by
\bea
	\vx_s & = & \vx + u_\parallel \hat{\v{n}} \, .
\eea
Using the fact that the galaxy density transforms as the 0-component of a 4-vector, we can derive the mapping up to third order, to obtain (see, for example, section 9.3.2 of \cite{Desjacques:2016bnm})
\begin{align}
	\delta_{g,s} =\:& \delta_g^\text{Jac} + \delta_g^\text{disp} \, , \nn \\
	\mbox{where}\quad
	\delta_g^\text{Jac} =\:& (1 + \delta_g) \left(1 - \eta_g + \eta_g^2\right) - \eta_g^3 -1 \, , \nn \\
	\delta_g^\text{disp} =\:& - u_{g\parallel} \partial_\parallel \delta_g^\text{Jac} + \frac12 u_{g\parallel}^2 \partial^2_\parallel 
	\delta_g^\text{Jac} + (u_{g\parallel} \partial_\parallel u_{g\parallel})\partial_\parallel \delta_g^\text{Jac} \, , \nn \\
        \eta_g =\:& \partial_\parallel u_{g\parallel}\,,
\label{eq:RSDmapping}
\end{align}
and all quantities are evaluated at the same apparent redshift-space spacetime point $(\vx_s,\tau)$. $\delta_g$ is the rest-frame galaxy density (but evaluated at the redshift-space position) containing the bias and selection contributions listed above. The mapping in \refeq{RSDmapping} can be expanded order by order, thus allowing for a consistent perturbative description of observed galaxy clustering, as derived in \cite{Scoccimarro:2004tg,Seljak:2011tx,Perko:2016puo,Desjacques:2018pfv}, to cite a few.

At linear order and on large scales, where $\eta_g = \eta$, \refeq{RSDmapping} simply yields $\delta_{g,s} = \delta_g - \eta$. If selection effects are absent, the $\eta$ contribution is not degenerate with the bias term in $\delta_g$, allowing for a direct constraint on the linear growth rate $f$. If, on the other hand, selection effects are present, then $\eta$ is multiplied by an additional free bias parameter $b_\eta$, so that we expect $b_\eta$ and $f$ to be degenerate at linear order (in the following, we will combine both RSDs and selection effects into combined effective coefficients so that $b_\eta=-1$ corresponds to the absence of selection effects). Fortunately, the contributions in $\delta_g^\text{disp}$, which start at second order, do not involve additional free bias parameters since they correspond to the displacement of the galaxy positions into redshift space which are directly controlled by the galaxy velocity; they are thus protected by the equivalence principle. We will see that the bispectrum in particular allows us to break the degeneracy between selection effects and RSDs via these displacement terms. This is analogous to how the real-space galaxy bispectrum allows breaking the $b_1-\sigma_8$ degeneracy; see, for example, \cite{Schmidt:2018bkr}.


\section{Observed galaxy statistics}
\label{sec:reviewgalaxy}

Given the galaxy density field in redshift space including selection effects, one can compute the galaxy power spectrum including its NLO or 1-loop correction, as well as the tree-level galaxy bispectrum. We refer the reader to \cite{Desjacques:2018pfv} for details of this calculation and merely reproduce the results here.

Let $\mu$, $\mu_i$ be the cosines of the angles between $\vk$, $\vk_i$ and the line-of-sight. The 1-loop galaxy power spectrum is given by
\bea
	P_{g,s}(k,\mu) & = & P_{g,s}^{\rm l+hd}(k,\mu) + P_{g,s}^{2-2}(k,\mu) + 2P_{g,s}^{1-3}(k,\mu) \, ,
\label{eq:pgsfull}
\eea
where
\bea
	P_{g,s}^{\rm l+hd}(k,\mu) & = & \( b_1 - b_{\eta} \mu^2 f \)^2 P_L(k) + P_{\epsilon}^{\{0\}} \nonumber \\
	& & \quad - \ 2 \Big\{ b_1 b_{\nabla^2\delta} - b_{\eta} \mu^2 f \( b_{\nabla^2\delta} + b_1 \bI + b_1 \bII \mu^2 \) \nonumber \\
	& & \qquad\quad + \ b_{\eta}^2 \mu^4 f^2 \( \bI + \bII \mu^2 \) \Big\} k^2 P_L(k) \nonumber \\
	& & \quad + \ k^2 P_{\eps}^{\{2\}} + b_{\eta} \mu^2 k^2 P_{\eps\varepsilon_{\eta}}^{\{2\}} \, .
\label{eq:pgslhd}
\eea
The 2-2 power spectrum reads
\bea
	P_{g,s}^{2-2}(k,\mu) & = & \sum_{n=0}^4 \sum_{(m,p)} A_{n(m,p)} (f,\{b_{O}\}_{2-2}) {\cal I}_{mp}(k) \mu^{2n} \, ,
\label{eq:pgs22}
\eea
with $\{b_{O}\}_{2-2} = \big\{ b_1, b_2, b_{K^2}, b_{\eta}, b_{\Pi_{\parallel}^{[2]}}, b_{(KK)_{\parallel}}, b_{\delta\eta}, b_{\eta^2} \big\}$ being the set of linear and second-order bias parameters. The functions $A_{nmp} (f,\{b_{O}\}_{2-2})$ are given in the {\tt Mathematica} supplement of \cite{Desjacques:2018pfv}. The 1-3 power spectrum is given by
\bea
	P_{g,s}^{1-3}(k,\mu) & = & \left\{ \sum_{l=0}^3 \sum_{n=1}^5 C_n^{1-3,\ell} (f,\{b_{O}\}_{1-3}) {\cal I}_{n}(k) {\cal L}_{2\ell}(\mu) \right\} P_L(k) \, ,
\label{eq:pgs13}
\eea
with $\{b_{O}\}_{1-3} = \big\{ b_1, b_{K^2}, b_{\rm td}, b_{\eta}, b_{\Pi_{\parallel}^{[2]}}, b_{(KK)_{\parallel}}, b_{\delta\eta}, b_{\eta^2}, b_{\delta \Pi_{\parallel}^{[2]}}, b_{\eta \Pi_{\parallel}^{[2]}}, b_{(\Pi^{[2]} K)_{\parallel}}, b_{\Pi_{\parallel}^{[3]}} \big\}$ being the set of bias parameters that yield nontrivial 1--3-type loop contributions. The functions $C_n^{1-3,\ell} (f,\{b_{O}\}_{1-3})$ are also given in the {\tt Mathematica} supplement of \cite{Desjacques:2018pfv}.

The tree-level galaxy bispectrum is given by
\bea
	& & B_{g,s}^{\rm LO}(k_1,\mu_1;k_2,\mu_2;k_3,\mu_3) = \bigg[ \frac{1}{3} B_{\epsilon}^{\{0\}} + 2\(b_1 - b_{\eta} \mu_2^2 f\) \(b_1 - b_{\eta} \mu_3^2 f\) \sum_{\{b_O\}_{2-2}} b_{O} S_{O} (\vk_2,\vk_3,f) \nonumber \\
	& & \qquad \times \ P_L(k_2) P_L(k_3) + 2\(b_1 - b_{\eta} \mu_1^2 f\) \( P_{\epsilon\epsilon_{\delta}}^{\{0\}} - \mu_1^2 f P_{\epsilon\epsilon_{\eta}}^{\{0\}} \) P_L(k_1) \bigg] + \, {\rm 2 \ perm.} \, ,
\label{eq:bgs}
\eea
where the functions $S_{O}(\vk_2,\vk_3,f)$ are given in table 1 of \cite{Desjacques:2018pfv}.


\section{Fisher analysis}
\label{sec:fisher}

The accuracy with which a given survey can measure cosmological parameters and the degree to which they are correlated can be estimated using the Fisher information matrix formalism. If the likelihood surface around the peak (maximum-likelihood point) can be approximated by a multivariate Gaussian distribution, then the Fisher matrix analysis allows us to compute the parameter covariance matrix. The likelihood function for a general cosmological parameter may of course not be Gaussian. Fisher forecasts are, therefore, not completely accurate and are, in such a case, only an approximation based on the curvature matrix (Hessian) at the peak. This issue can be resolved by probing the full parameter space using, for example, Markov-Chain Monte-Carlo (MCMC) sampling methods. Nevertheless, the Fisher matrix formalism provides a good estimate of parameter uncertainties and correlations, and has been successfully employed in examining the statistical contents of the CMB and LSS probes. In this section we apply it to forecast constraints on cosmological parameters with observed galaxy clustering.

The Fisher matrix is defined as the Hessian matrix of the log-likelihood, ${\cal L} \equiv \ln L$, of galaxy clustering data in the parameter space $\boldsymbol{\theta}$,
\bea
	F_{ab} & = & - \left\langle \frac{\partial^2 {\cal L}}{\partial \theta^a \partial \theta^b} \right\rangle .
\eea
Using the Cram\'{e}r-Rao bound, the inverse matrix, $(F^{-1})_{ab}$, yields an estimate of the best possible covariance matrix (that is, the minimum uncertainty) for measurement errors on the parameters.

Regarding the choice of cosmological parameters, there are two paths that one can follow. The first is to consider a parameter space of cosmological models, such as smooth dark energy scenarios described by a time-varying equation of state $w_{\rm de}(z)$ parametrized in some form. The alternative approach is to remain more model-independent and choose parameters that are closer to the data, such as the angular diameter distance $D_A(z)$, Hubble rate $H(z)$, and logarithmic growth rate $f(z)$ at some effective redshift $z$. These parameters have traditionally been used in the analysis of galaxy redshift surveys, since the baryon acoustic oscillation (BAO) feature and Alcock-Paczynski (AP) distortions allow for a fairly model-independent measurement of $D_A(z)$ and $H(z)$, while large-scale RSDs yield a measurement of $f^2(z)$ times the matter power spectrum; the amplitude of the latter is often parametrized through the r.m.s. variation $\sigma_8$ of density fluctuations smoothed with a spherical-tophat filter of radius $8 \, h^{-1} \, {\rm Mpc}$, so that RSDs constrain the combination $f(z) \sigma_8(z)$.  From here, we will suppress any redshift dependence of cosmological parameters to avoid clutter.

In this paper we follow the second approach, with $D_A$, $H$, and $f$ forming the first three cosmological parameters in our analysis. In order for the constraints on these to be independent of the cosmological model, we also allow for freedom in the matter power spectrum normalization and shape, by adding $\Omega_{c0}$, the CDM density parameter today, and $n_s$ and $n_{\rm run}$, the tilt and running of the power spectrum of primordial scalar perturbations, to our set of parameters. We use flat $\Lambda$CDM as the background cosmology model, so varying $\Omega_{c0}$ corresponds to changing the cosmological constant accordingly. While we keep the normalization $A_s$ of primordial curvature perturbations fixed, the amplitude of the late-time linear power spectrum varies with $\Omega_{c0}$ through the linear growth factor. We will forecast the constraints on our six cosmological parameters $\{D_A,H,f,\Omega_{c0},n_s,n_{\rm run}\}$ after marginalizing over all bias parameters listed in section\ \ref{sec:reviewgalaxy} that describe galaxy clustering, specifically the power spectrum at 1-loop order and bispectrum at tree-level. The full set of parameters $\boldsymbol{\theta}$ considered here can be found in table\ \ref{table:euclid3}.

In the subsections below, we detail the calculation of the Fisher matrix for the galaxy power spectrum and bispectrum. We also present details for estimating the systematic bias arising from fixing a model parameter to the wrong value, or from theoretical systematics in the modeling of the power spectrum and bispectrum.

\subsection{Galaxy power spectrum contribution}

In the continuum limit, the Fisher matrix for the galaxy power spectrum can be approximated as \cite{Tegmark:1997rp}
\bea
    F_{ab} & = & 2\pi \int_0^{k_{\rm max}} \d \ln k \int_{-1}^1 \frac{\d \mu}{2} \, \frac{\partial \ln P_g(k,\mu)}{\partial \theta^a} \frac{\partial \ln P_g(k,\mu)}{\partial \theta^b} w(k,\mu) \, ,
\label{eq:pkfisher}
\eea
where $P_g(k,\mu)$ denotes the galaxy power spectrum in eq.\ (\ref{eq:pgsfull}) and the weight function $w(k,\mu)$ is defined as
\bea
	w(k,\mu) & = & \frac{k^3}{(2\pi)^3} V_{\rm survey} \, ,
\label{eq:weightfn}
\eea
with $V_{\rm survey}$ being the survey volume. Note that, in our convention, $P_g$ includes the shot-noise contribution via the parameter $P_{\eps}^{\{0\}}$. Hence, $w(k,\mu)$ simply provides the mode count and does not contain any noise.

In appendix \ref{app:pkderivatives} we calculate the logarithmic derivatives of $P_g(k,\mu)$ with the six cosmological parameters we consider. When calculating the numerical derivatives, we use the finite-difference method with step sizes of order of the final parameter constraints.

\subsection{Galaxy bispectrum contribution}

For the galaxy bispectrum, the Fisher matrix is given by \cite{Desjacques:2016bnm}
\bea
	F_{ab} & = & \sum_{(k_1,k_2,k_3)} \int_{-1}^{1} d\mu \int_0^{2\pi} d\phi \,\frac{1}{s_B V_{\rm survey}} \frac{1} {P_g(k_1,\mu_1)P_g(k_2,\mu_2)P_g(k_3,\mu_3)} \nonumber \\
	& & \qquad \quad \times \, \frac{\partial B_g(\vk_1,\vk_2,\vk_3)}{\partial \theta^a} \frac{\partial B_g(\vk_1,\vk_2,\vk_3)}{\partial \theta^b} \( \prod_{i=1}^3 \frac{k_i\Delta k_i}{k_{F_i}^2} \) \times
	\left\{
	    \begin{array}{cc}
        		\pi, & k_1 = k_2 + k_3\\
	        2\pi, & {\rm otherwise}
	    \end{array}
	\right. , \quad
\label{eq:bkfisher}
\eea
where $B_g(\vk_1,\vk_2,\vk_3)$ denotes the galaxy bispectrum in eq.\ (\ref{eq:bgs}). Here, $s_B$ is a symmetry factor (6 for equilateral triangles, 2 for isosceles triangles, and 1 for other triangles), and $\Delta k_i$ and $k_{F_i} = 2\pi/L_i$  are, respectively, the Fourier-space bin size and the fundamental wavenumber in the $i^{\rm th}$ direction. For simplicity, we approximate the survey volume as a cube: $L_i=V_{\rm survey}^{1/3}$. Also, $\mu$ here is the cosine of the angle between the line-of-sight direction and the plane embedding $(\vk_1,\vk_2,\vk_3)$, while $\phi$ is the angle between the line-of-sight direction projected onto the plane and $\vk_1$. Further defining $\alpha$ as the inner angle between $\vk_1$ and $\vk_2$, the parallel components of all three vectors are given by $k_{1\parallel} = k_1 \mu \cos\phi$, $k_{2\parallel} = k_2 \mu \cos(\alpha+\phi)$, and $k_{3\parallel} = -k_{1\parallel}-k_{2\parallel}$.

Note that only the triplets $(k_1,k_2,k_3$) forming a triangle contribute to the integral in eq.\ (\ref{eq:bkfisher}); we ensure this by checking that the triangle inequality holds for the magnitudes $(k_1, k_2, k_3)$. Once we have ensured that $(k_1,k_2,k_3)$ correspond to a triangle, we compute $\alpha$ and, thereby, $k_{2\parallel}$ and $k_{3\parallel}$ needed for $B_g^\text{LO}$. We also use the following simplifications to speed-up the calculation of $F_{ab}$ for the bispectrum. First, since the integrand is symmetric under all permutations of $(\vk_1,\vk_2,\vk_3)$, we restrict the integral limits on the magnitudes to $0 \le k_3 \le k_2 \le k_1 \le k_{\rm max}$, inserting a factor of 6. Second, we exploit the symmetry under $\mu \rightarrow -\mu$ and $\phi \rightarrow \phi + \pi$, to change the limits on the $\mu$ integral to $0$ to $1$, multiplying the resulting integral by a factor of $2$. With these simplifications the total factor multiplying the resulting integrals is $12$.

Appendix \ref{app:bkderivatives} details the calculation of the derivatives of $B_g(\vk_1,\vk_2,\vk_3)$ for the six cosmological parameters we consider.

\subsection{Parameter shifts and theoretical systematics}

If one or several parameters not marginalized over in a likelihood analysis are set to incorrect values, then the best-fit value of the remaining parameters can be systematically biased. Systematic errors also arise when the theoretical model has a limited range of validity, which is always the case in perturbation theory. Namely, we consider in this paper the galaxy power spectrum at 1-loop and bispectrum at tree-level. Ignoring the next higher order, 2-loop (1-loop) contributions to the galaxy power spectrum (bispectrum), can also give rise to parameter shifts. The systematic bias caused by these sources can be estimated with the Fisher matrix approach provided that the resulting systematic errors do not significantly exceed the statistical error; see, for instance, \cite{Dodelson:2005ir,Heavens:2009nx,Gebhardt:2018zuj}.

To proceed, we write the full parameter vector as $\boldsymbol{\theta}=(\boldsymbol{\alpha},\boldsymbol{\beta})$, where $\boldsymbol{\beta}$ denotes the parameters that are fixed to incorrect values during the cosmological parameter estimation. The vector $\boldsymbol{\beta}$ includes, among others, coefficients that multiply the next-order perturbative contributions. These coefficients are set to unity (zero) when the next-order perturbative contributions are (are not) included. Another example is the selection bias parameter that is ignored in the conventional analysis of the galaxy surveys to date.

We can model the maximum-likelihood estimator for cosmological parameters $\boldsymbol{\alpha}$ as maximizing the function
\be
	\mathcal L(\boldsymbol{\alpha},\boldsymbol{\beta}) + \boldsymbol{\lambda}\cdot\big(\boldsymbol{\beta}-\boldsymbol{\beta}_f\big) \, ,
\ee
where $\mathcal L\equiv \ln L$ is the logarithm of the likelihood function $L$ for the full model, $\boldsymbol{\lambda}$ are Lagrange multipliers, and $\boldsymbol{\beta}_f$ are false values of the parameters $\boldsymbol{\beta}$. Varying with respect to $\boldsymbol{\alpha}$, $\boldsymbol{\beta}$, and $\boldsymbol{\lambda}$ gives us three equations to solve for the maximum,
\bea
	\frac{\partial\mathcal L}{\partial\boldsymbol{\alpha}}(\boldsymbol{\alpha}_c,\boldsymbol{\beta}_c) & = & 0 \, , \nn \\
	\frac{\partial\mathcal L}{\partial\boldsymbol{\beta}}(\boldsymbol{\alpha}_c,\boldsymbol{\beta}_c) + \boldsymbol{\lambda}_c & = & 0 \, , \nn \\
	\boldsymbol{\beta}_c - \boldsymbol{\beta}_f & = & 0 \, , \label{eq:lagrange}
\eea
where $(\boldsymbol{\alpha}_c,\boldsymbol{\beta}_c)$ designates the position of the constrained maximum likelihood, which is shifted from the true position $(\boldsymbol{\alpha}_t,\boldsymbol{\beta}_t)$ according to
\bea
	(\boldsymbol{\alpha}_c,\boldsymbol{\beta}_c) & = & (\boldsymbol{\alpha}_t+\Delta\boldsymbol{\alpha},\boldsymbol{\beta}_t+\Delta\boldsymbol{\beta}) \, .
\eea
Only the first and third equations of \refeq{lagrange} are useful to our purpose (the knowledge of $\boldsymbol{\lambda}_c$ is necessary solely for the computation of $\Delta\mathcal L$). Expanding the first equation around the true maximum gives
\bea
	\frac{\partial^2\mathcal L}{\partial\boldsymbol{\alpha}^2}(\boldsymbol{\alpha}_t,\boldsymbol{\beta}_t)\,\Delta\boldsymbol{\alpha}+\frac{\partial^2\mathcal L}{\partial\boldsymbol{\alpha}\partial\boldsymbol{\beta}}(\boldsymbol{\alpha}_t,\boldsymbol{\beta}_t)\,\Delta\boldsymbol{\beta} & = & 0
\eea
at first order in the parameter shifts $(\Delta\boldsymbol{\alpha},\Delta\boldsymbol{\beta})$. Since the third equation of \refeq{lagrange} tells us that $\Delta\boldsymbol{\beta}=\boldsymbol{\beta}_f-\boldsymbol{\beta}_t$, this can also be written as
\bea
	\Delta\boldsymbol{\alpha} & = & - \left(\frac{\partial^2\mathcal L}{\partial\boldsymbol{\alpha}^2}\right)^{-1} \left(\frac{\partial^2\mathcal L}{\partial\boldsymbol{\alpha}\partial\boldsymbol{\beta}}\right)\big(\boldsymbol{\beta}_f-\boldsymbol{\beta}_t\big) \, .
\eea
Note that this relation remains valid to first order in the parameter shifts, whether the partial derivatives are evaluated at the true maximum or at the constrained position. Upon defining $F'$ as the Fisher matrix of the reduced model comprised of the parameters $\boldsymbol{\alpha}$ only, we can express this result in the more familiar form,
\bea
	\Delta\alpha^a & = & - \big(F^{' -1}\big)^a{}_b G^b{}_c \, \Delta \beta^c \, .
\label{eq:shifts}
\eea
The matrix $G$ has entries
\bea
	G_{ab} & = & -\frac{\partial^2 \mathcal L}{\partial\alpha^a \partial\beta^b}(\boldsymbol{\alpha}_t,\boldsymbol{\beta}_t) \ \simeq \ -\frac{\partial^2 \mathcal L}{\partial\alpha^a \partial\beta^b}(\boldsymbol{\alpha}_f,\boldsymbol{\beta}_f) \, ,
\eea
where again the second equality holds at first order in parameter shifts.

\subsubsection{Model systematics}
\label{sec:shiftsparam}

Consider first a generic parameter error $\Delta\beta^a$ in the fiducial statistics. The resulting parameter shifts $\Delta\alpha^a$ are all computed from \refeq{shifts}, but the computation of $G_{ab}$ differs among the statistics. For the 1-loop galaxy power spectrum, the vector $G^a{}_b \Delta\beta^b$ is obtained upon replacing in \refeq{pkfisher} the factor $\frac{\partial \ln P_g(k,\mu)}{\partial \theta^b}$ with $\frac{\partial \ln P_g(k,\mu)}{\partial \beta^b} \Delta\beta^b$. For the tree-level galaxy bispectrum, $G^a{}_b \Delta\beta^b$ is obtained by replacing in \refeq{bkfisher} the factor $\frac{\partial B_g(\vk_1,\vk_2,\vk_3)}{\partial \theta^b}$ with $\frac{\partial B_g(\vk_1,\vk_2,\vk_3)}{\partial \beta^b} \Delta\beta^b$.

We have focused here on systematic shifts produced by fixing one parameter (of the extended model) to an incorrect value. In principle, several parameters could be assigned incorrect values, and the systematic shift resulting from the combination of this assignment could be quite different. We will not consider such a possibility here.

\subsubsection{Theory systematics}
\label{sec:shiftspkbk}

Consider now a higher-order contribution $P_{g,\rm sys}(k,\mu)$ to the galaxy power spectrum. Taking advantage of the similarity between $G_{ab}$ and $F_{ab}$, the parameter shifts arising from ignoring this term can also be written as in eq.\ (\ref{eq:shifts}), with the vector $G^a{}_b \Delta\beta^b$ obtained by replacing in \refeq{pkfisher} the factor $\frac{\partial \ln P_g(k,\mu)}{\partial \theta^b}$ by $\frac{P_{g,{\rm sys}}(k,\mu)}{P_g(k,\mu)}$,
\bea
    \Delta\alpha^a & = & 2\pi \big(F^{' -1}\big)^a{}_b \int_0^{k_{\rm max}} \d \ln k \int_{-1}^1 \frac{\d \mu}{2} \, \frac{\partial \ln P_g(k,\mu)}{\partial \theta_b}\frac{P_{g,{\rm sys}}(k,\mu)}{P_g(k,\mu)} \, w(k,\mu) \, .
\label{eq:gijdbj}
\eea
Note that $\beta_f = 0$ while $\beta_t=1$. Furthermore, $F'$ is identical to $F$ here since we are not really considering a reduced parameter space.

We consider two neglected higher-order contributions, namely the contribution from the 2-loop matter power spectrum and higher-derivative bias,
\bea
	P_{g,\rm sys}(k,\mu) & = & (b_1 + \mu^2 f)^2 P_{m,{\rm 2-loop}}(k) - 2(b_1 + \mu^2 f) k^4 R_*^4 \(1 + k^2 R_*^2\) P_L(k) \, ,
\label{eq:pgsys}
\eea
where $R_*$ is the scale controlling the higher-derivative bias parameters. We will take $R_* = 1 \, h^{-1} \, {\rm Mpc}$ here. This scale equals the Lagrangian radius of halos with mass $\sim 5\times 10^{11} M_\odot$ for which the observed ratio of stellar mass to halo mass peaks.  The existence of such a peak in the ratio of stellar and halo masses is presumed to result from the efficiency of supernovae and AGN feedback at low and high halo masses, respectively (see \cite{GalaxyFormationBook} and references therein). Interestingly, this characteristic halo mass scale appears to vary weakly across a wide range of redshift; see, for example, \cite{Behroozi:2012iw}. In fact, it turns out to be close to the median mass of the dark matter halos hosting the {\small H}$\alpha$ and {\small Ly}$\alpha$ emitters to be surveyed by Euclid, the Roman Space Telescope, or HETDEX \cite{Pozzetti:2016cch,Merson:2017efv,khostovan/etal:2019}. Therefore, it is a natural choice for $R_*$. Also, we have restricted to $\mu^4$ terms in eq.\ (\ref{eq:pgslhd}) whereas the true 2-loop contribution will contain higher powers of $\mu$. This restriction is motivated by the fact the tree-level power spectrum in eq.\ (\ref{eq:pgslhd}) goes up to $\mu^4$, and we expect it to contain most of the cosmological information. We will later use the error arising from ignoring the contribution in eq.\ (\ref{eq:pgsys}) to determine the value of $k_{\rm max}$ for our analysis. It is also possible to choose different $k_{\rm max}$ values for different multipoles as studied, for example, in \cite{Markovic:2019sva}.
 
Similarly,  parameter shifts due to a higher-order contribution $B_{g,\rm sys}(\vk_1,\vk_2,\vk_3)$ to the galaxy bispectrum are given by eq.\ (\ref{eq:shifts}), with the vector $G^a{}_b \Delta\beta^b$ obtained by replacing in eq.\ (\ref{eq:bkfisher}) the factor $\frac{\partial B_g(\vk_1,\vk_2,\vk_3)}{\partial \theta^b}$ with $B_{g,{\rm sys}}(\vk_1,\vk_2,\vk_3)$. In case of the bispectrum as well we consider two neglected contributions, namely that from the 1-loop matter bispectrum and the leading higher-derivative bias contribution,
\ba
	B_{g,\rm sys}(\vk_1,\vk_2,\vk_3) & = & (b_1 + \mu_1^2 f) (b_1 + \mu_2^2 f) (b_1 + \mu_3^2 f) B_{m,{\rm 1-loop}}(k_1,k_2,k_3) \nn \\
	& & + \left[(b_1 + \mu_1^2 f) (b_1 + \mu_2^2 f) k_3^2 R_*^2 + 2\,{\rm perm.} \right] B_{m,{\rm tree}}(k_1,k_2,k_3) \, .
\ea

In the next section we will evaluate the parameter shifts arising from these theory systematics in order to determine the optimal $k_{\rm max}$ values for our power spectrum and bispectrum analysis.


\section{Results}
\label{sec:fisherresults}

We now present the results of our Fisher matrix analysis. We forecast constraints on the six cosmological parameters $\{D_A, H, f, \Omega_{\rm c0}, n_s, n_{\rm run}\}$ for three galaxy surveys: Euclid, the Roman Space Telescope, and HETDEX. For the fiducial cosmology, we use the flat $\Lambda$CDM parameter values in the {\tt base\_plikHM\_TTTEEE\_lowTEB\_lensing\_post\_BAO\_H080p6\_JLA} column of Planck 2015 \cite{Adam:2015rua,Ade:2015xua}: $\Omega_{b0}h^2 = 0.022307$, $\Omega_{c0}h^2 = 0.11865$, $\Omega_{\nu 0}h^2 = 0.000638$, $\Omega_{\Lambda0} = 0.69179$, $h = 0.6778$, $n_s = 0.9672$, ${\cal A}_s = 2.147 \times 10^{-9}$, and $w = -1$. To facilitate the reading of this section, we have grouped the tables containing detailed forecasts in appendix\ \ref{app:tableresults}.

As for the galaxy bias, we consider two scenarios depending on whether selection effects are included or not, namely,
\begin{itemize}
	\item \emph{With selection effects}: The bias parameters
\be
	b_{\eta}, \ \bII, \ b_{\Pi_{\parallel}^{[2]}},\ b_{(KK)_{\parallel}}, \
	b_{\delta\eta}, \ b_{\eta^2},\ b_{\delta \Pi_{\parallel}^{[2]}}, \
	b_{\eta \Pi_{\parallel}^{[2]}}, \ b_{(\Pi^{[2]} K)_{\parallel}}
\ee
are allowed to vary and are marginalized over (as discussed below, we do not marginalize over $b_{\Pi_{\parallel}^{[3]}}$). A summary of parameters and their fiducial values is given in tables\ \ref{table:euclid1} and \ref{table:euclid3}.
	\item \emph{Without selection effects}: All parameters mentioned above are fixed to their fiducial value, in particular $b_{\eta} = -1$ and $b_{\delta\eta} = -b_1$. A summary of the remaining parameters and their fiducial values is given in tables\ \ref{table:euclid1} and \ref{table:euclid2}.
\end{itemize}
All rest-frame (non-selection) deterministic biases and stochastic amplitudes are always margi- nalized over. In terms of cosmology, we likewise consider two scenarios,
\begin{itemize}
	\item The full set of six cosmological parameters $\{D_A, H, f, \Omega_{\rm c0}, n_s, n_{\rm run}\}$, which we constrain with the power spectrum alone and a combination of the power spectrum and bispectrum. Note that the power spectrum alone only yields very weak constraints on $f$, because it is sensitive mostly to the parameter combination $f \sigma_8$, and $\sigma_8$ depends on $\Omega_{c0}$, $n_s$, and $n_{\rm run}$. This degeneracy is, however, broken when the bispectrum is included (see, for example, \cite{Heavens:1998es,Scoccimarro:1999ed,Gil-Marin:2014pva}).
	\item A reduced set of three cosmological parameters $\{D_A, H, f\}$, keeping $\Omega_{c0}$, $n_s$, and $n_{\rm run}$ constant, constrained with the power spectrum alone. The parameter constraints in this scenario correspond to the conventional analysis of measuring $\{D_A, H, f\sigma_8\}$ using the combination of BAO, AP, and RSDs with the galaxy power spectrum.
\end{itemize}

\subsection{Euclid}

\begin{figure}[!t]
\begin{center}
	\includegraphics[width=2.85in,angle=0]{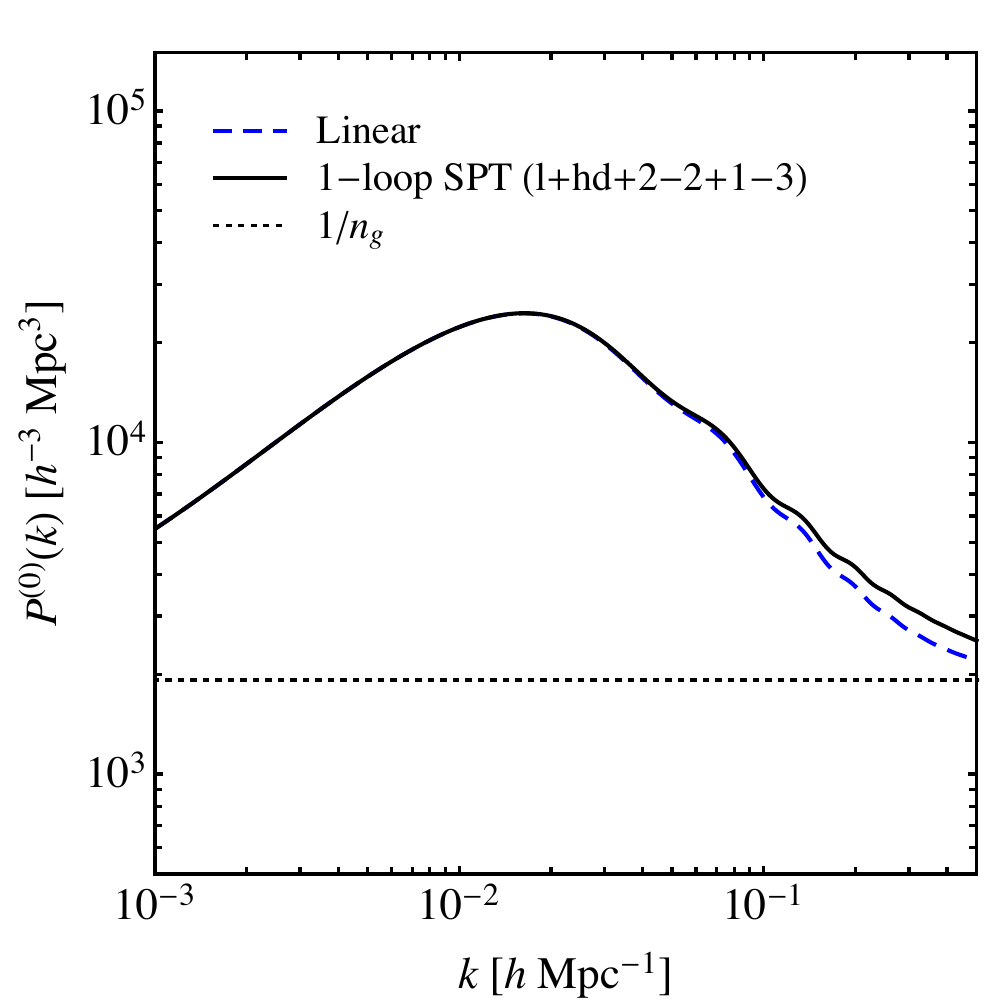}
	\includegraphics[width=2.85in,angle=0]{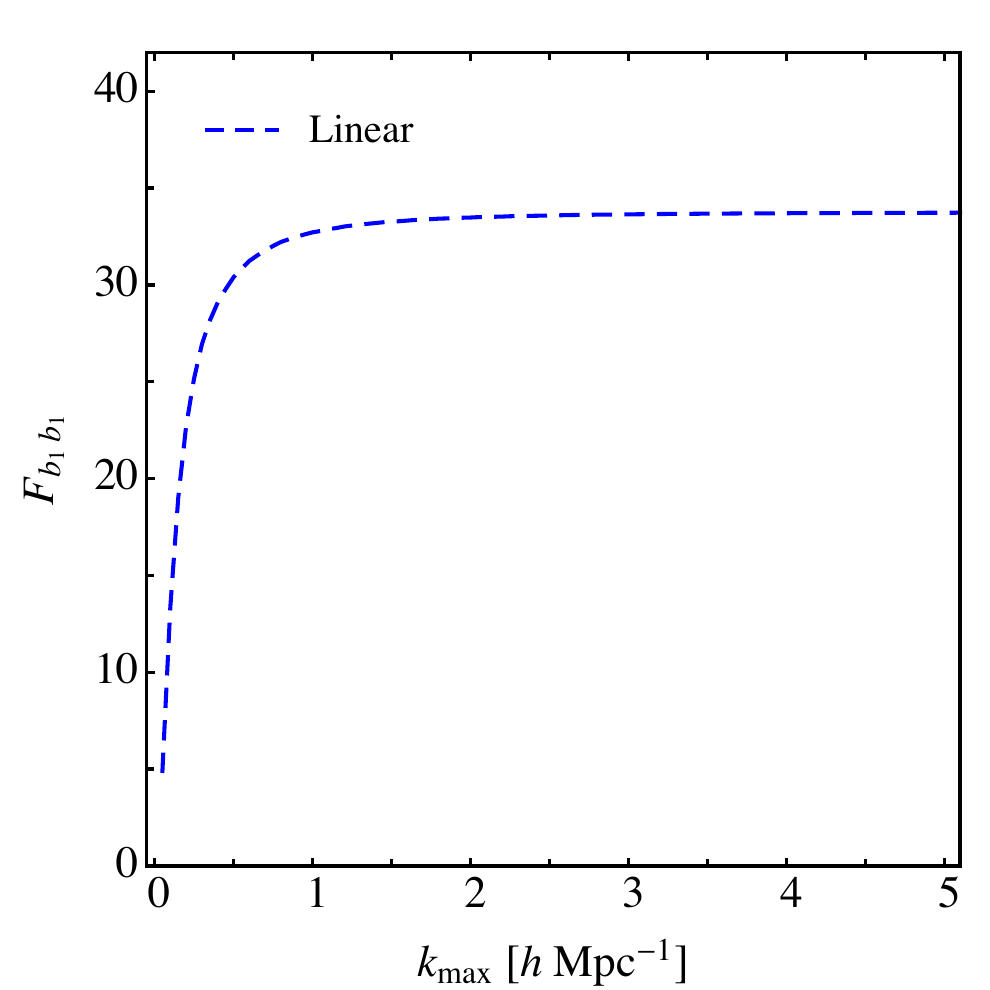}
	\caption{(Left) Monopole of the galaxy power spectrum at $z = 1.4$ using fiducial bias parameters. The dotted horizontal line shows the expectd shot noise for a Euclid-like survey. Note that the fiducial value of $P_{\epsilon}^{\{0\}}$ is $1/n_g$. (Right) Unmarginalized cumulative Fisher information $F_{b_1b_1}$ in the linear bias parameter $b_1$ as a function of $k_{\rm max}$ for a Euclid-like survey.
}
\label{fig:pkeuclid}
\end{center}
\end{figure}

For a Euclid-like survey, we use parameter values
\be
	z \, = \, 1.4 \, , \quad V_{\rm survey} \, = \, 63 \, h^{-3} \, {\rm Gpc}^3 \, , \quad n_g \, = \, 5.2 \times 10^{-4} \, h^3 \, {\rm Mpc}^{-3} \, , \quad b_1 \, = \, 1.5 \, .
\ee
In the left panel of fig.\ \ref{fig:pkeuclid}, we show the monopole of the linear and nonlinear galaxy power spectrum for our fiducial parameter choices.
As a rough indication of the information content as a function of scale, the right panel shows the unmarginalized cumulative Fisher information in the linear bias as a function of $k_{\rm max}$. The Fisher information increases with $k_{\rm max}$ on large scales, as the inclusion of more modes reduces cosmic variance. On small scales, on the other hand, the Fisher information saturates due to the sparsity of the galaxy sample (shot-noise). Note that the raw Fisher information is far from being saturated on the scales $k < k_{\rm max}$ where our perturbative prediction for the galaxy power spectrum is trustable. We turn to determining this $k_{\rm max}$ next.

\begin{figure}[!t]
\begin{center}
	\includegraphics[width=2.85in,angle=0]{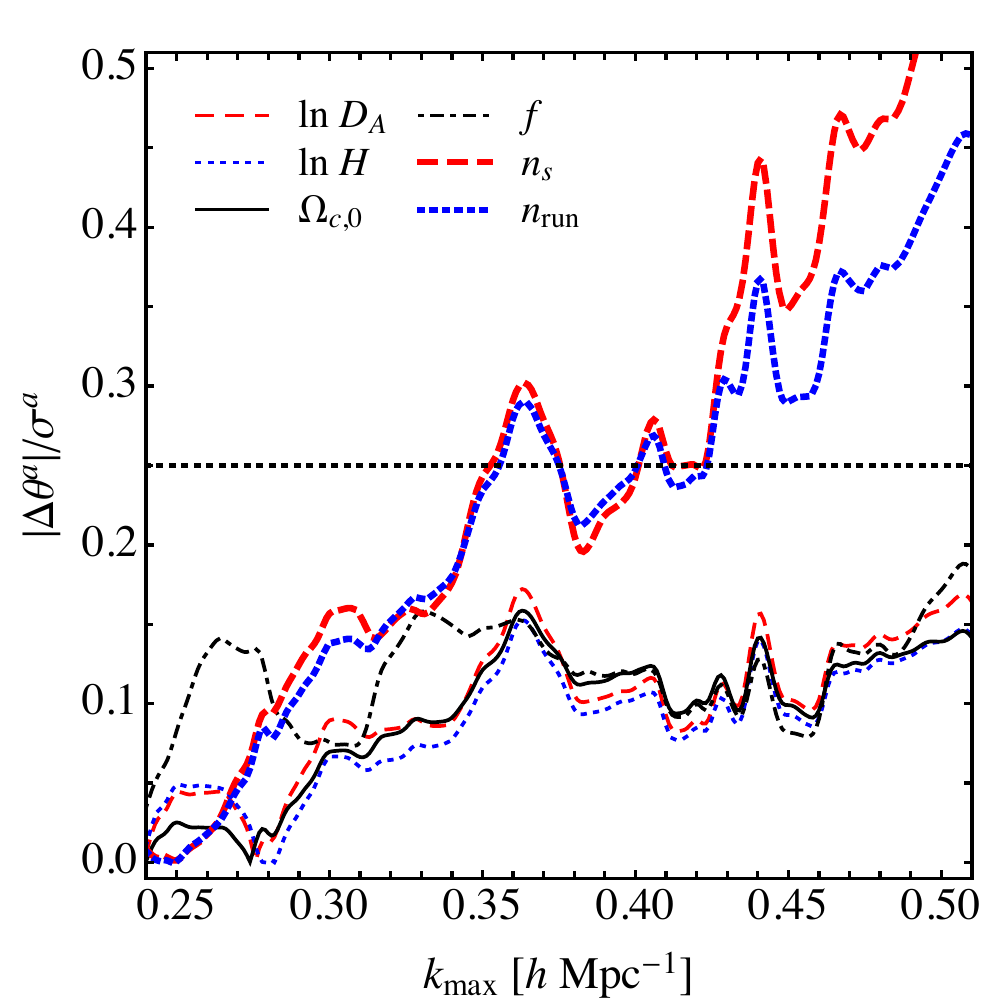}
	\includegraphics[width=2.85in,angle=0]{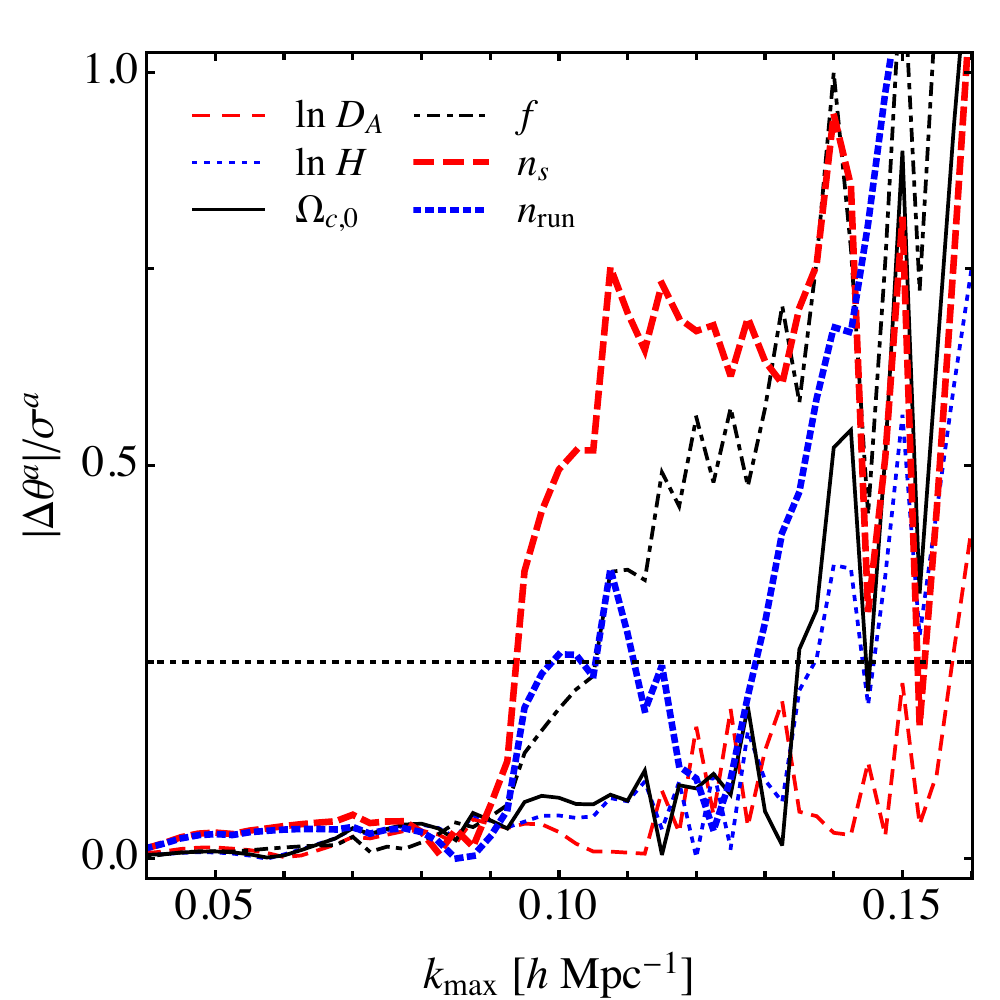}
	\caption{(Left) Parameter shifts, compared to the corresponding $1\sigma$ errors, for a Euclid-like survey when the 2-loop and higher-derivative bias contributions to the galaxy power spectrum are treated as theoretical systematics; here we use the power spectrum alone to different $k_{\rm max}$ values and include selection effects. (Right) Parameter shifts, compared to the corresponding $1\sigma$ errors, when the 1-loop and higher-derivative bias contributions to the galaxy bispectrum are treated as theoretical systematics; here we use the power spectrum up to $k_{\rm max} = 0.35 \, h \, {\rm Mpc}^{-1}$ with the bispectrum up to different $k_{\rm max}$ values and include selection effects. The horizontal lines at $|\Delta\boldsymbol{\theta}| = 0.25\boldsymbol{\sigma}$ show the maximum shift allowed.}
\label{fig:thetaieuclid}
\end{center}
\end{figure}

In the Fisher analysis that follows, we choose $k_{\rm max}$ separately for the galaxy power spectrum and bispectrum, such that theoretical systematics in each lead to shifts of $\lesssim 0.25\sigma$ in all six cosmological parameters. We determine the shifts using eq.\ (\ref{eq:shifts}) and the method described in section\ \ref{sec:shiftspkbk}. In practice, we first consider parameter shifts due to the 2-loop and higher-derivative bias contributions to the power spectrum. We calculate the Fisher matrix at the fiducial values shown in tables\ \ref{table:euclid2} and \ref{table:euclid3}, marginalizing over all bias and selection parameters, except for the parameter $b_{\Pi_{\parallel}^{[3]}}$ for reasons explained below. The shifts at different $k_{\rm max}$ values are shown in the left panel of fig.\ \ref{fig:thetaieuclid} and our criterion gives $k_{\rm max} = 0.35 \, h \, {\rm Mpc}^{-1}$ at $z=1.4$. We then adopt this value for the power spectrum and consider parameter shifts when the 1-loop and higher-derivative bias contributions to the bispectrum are treated as theoretical systematics. The resulting shifts are shown in the right panel of fig.\ \ref{fig:thetaieuclid}; these are noisier than those for the power spectrum and we choose a conservative value of $k_{\rm max} = 0.10 \, h \, {\rm Mpc}^{-1}$ for the bispectrum. We emphasize that, while they are determined in a systematic way, these choices of maximum wavenumber are still to be seen as very rough approximations, as our estimation only includes a part of the higher-order contributions. Taking into account the full higher-order contributions would require assumptions on the values of the higher-order bias parameters for the galaxy sample, however.

\subsubsection{Cosmology constraints}

\begin{figure}[!t]
\begin{center}
	\includegraphics[width=4in,angle=0]{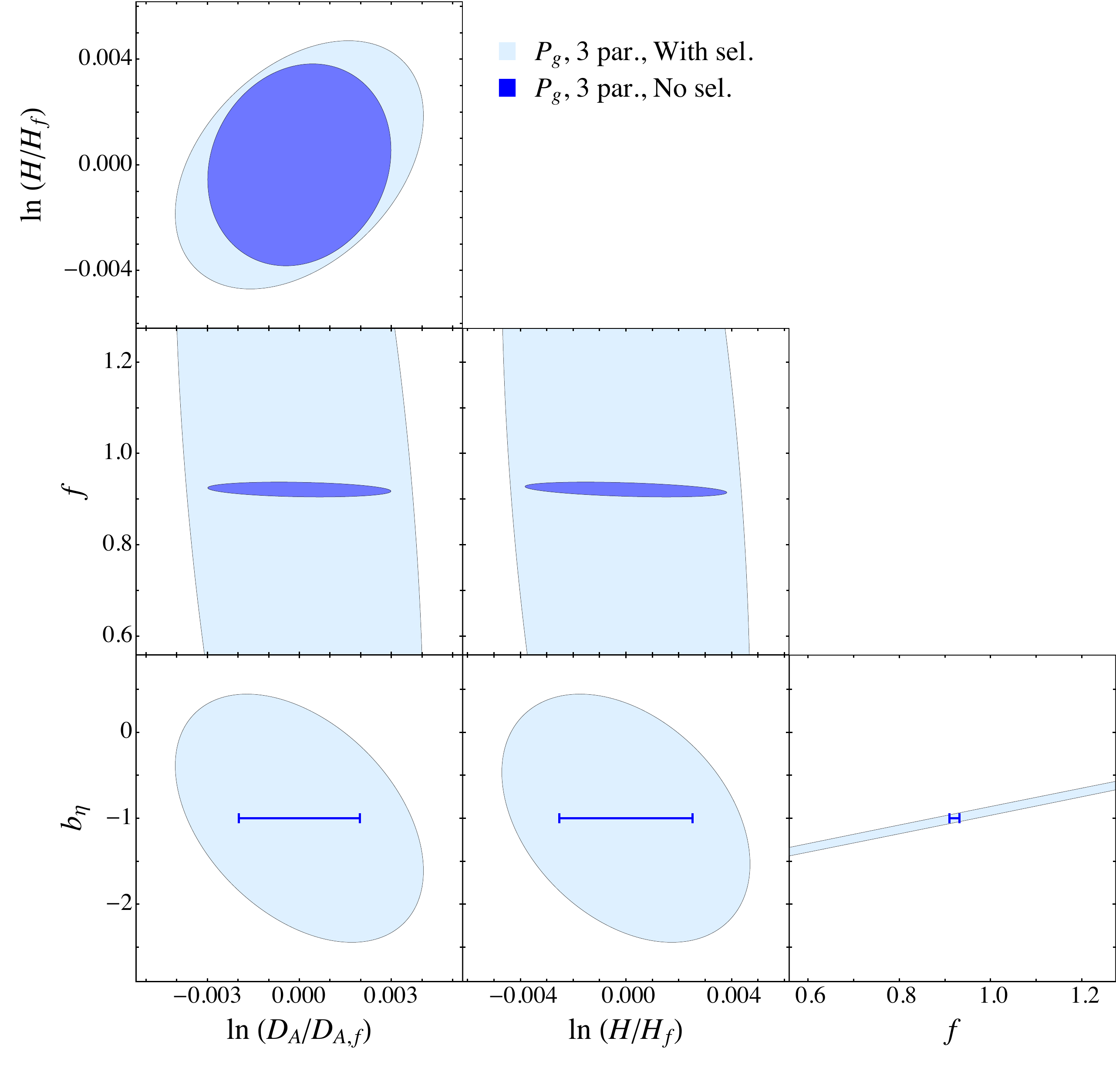}
	\caption{2D contour plots of the projected $1\sigma$ errors for a Euclid-like survey obtained using the power spectrum alone and varying only three of the six cosmological parameters. The bigger light blue ellipses include selection effects while the smaller dark blue ellipses ignore them. In the panels that show $b_\eta$, the horizontal $1\sigma$ error bars indicate the case where selection effects are ignored, with $b_\eta$ fixed to $-1$.}
\label{fig:triangleeuclid1}
\end{center}
\end{figure}

Having determined $k_\text{max}$, we can obtain error estimates on the various bias and cosmology parameters. We first consider constraints using only the power spectrum. As discussed above, the power spectrum alone is not sufficient to break the $f-\sigma_8$ degeneracy. We thus consider only three of the six cosmological parameters for now: $D_A$, $H$, and $f$; keeping $\Omega_{c0}$, $n_s$, and $n_{\rm run}$ constant uniquely determines $\sigma_8$. At $z=1.4$ and for the cosmological parameters we consider, the linear matter power spectrum gives $\sigma_8 = 0.422$.\footnote{In the spirit of using units of Mpc rather than $h^{-1} \, {\rm Mpc}$, as recently suggested in \cite{Sanchez:2020vvb}, we also quote the value $\sigma_{12} = 0.417$, where $\sigma_{12}$ is the r.m.s. density variation when smoothed with a spherical-tophat filter of radius $12 \, {\rm Mpc}$.} Although we could adopt a slightly higher value of $k_{\rm max}$ than was inferred from fig.\ \ref{fig:thetaieuclid} since the parameter space is reduced in the present case, we will stick to the conservative estimate $k_\text{max}=0.35 \, h \, {\rm Mpc}^{-1}$. Furthermore, we set the bias parameters to the fiducial values given in table\ \ref{table:euclid1}, which also displays the forecasted $1\sigma$ errors when selection effects are ignored or included. In the presence of selection effects, the bias parameters $b_{\rm td}$, $b_{\delta \Pi_{\parallel}^{[2]}}$, $b_{(\Pi^{[2]} K)_{\parallel}}$, and $b_{\Pi_{\parallel}^{[3]}}$ are highly degenerate with one another, with correlations reaching up to $r = \pm 1$. We therefore drop one of the parameters, $b_{\Pi_{\parallel}^{[3]}}$, from our analysis. Furthermore, the growth rate $f$ is strongly correlated (which we define as correlation coefficient $|r| \gtrsim 0.9$) with various parameters. In particular, we find $r \approx 1.0$ with $b_{\eta}$, which is as expected following our discussion above; further, $r = 0.95$ with $b_{K^2}$, $-0.91$ with $b_{\rm td}$, $-0.95$ with $b_{(KK)_{\parallel}}$, and $0.94$ with $b_{(\Pi^{[2]} K)_{\parallel}}$. By contrast, in the absence of selection effects, $f$ is only weakly degenerate (which we define as $|r| \lesssim 0.9$) with any other parameter. We also show in fig.\ \ref{fig:triangleeuclid1} the 2D contour plots for the three cosmological parameters and $b_{\eta}$ in these two cases. Note that the subscripts $f$ here indicate fiducial values.

\begin{figure}[!t]
\begin{center}
	\includegraphics[width=6.0in,angle=0]{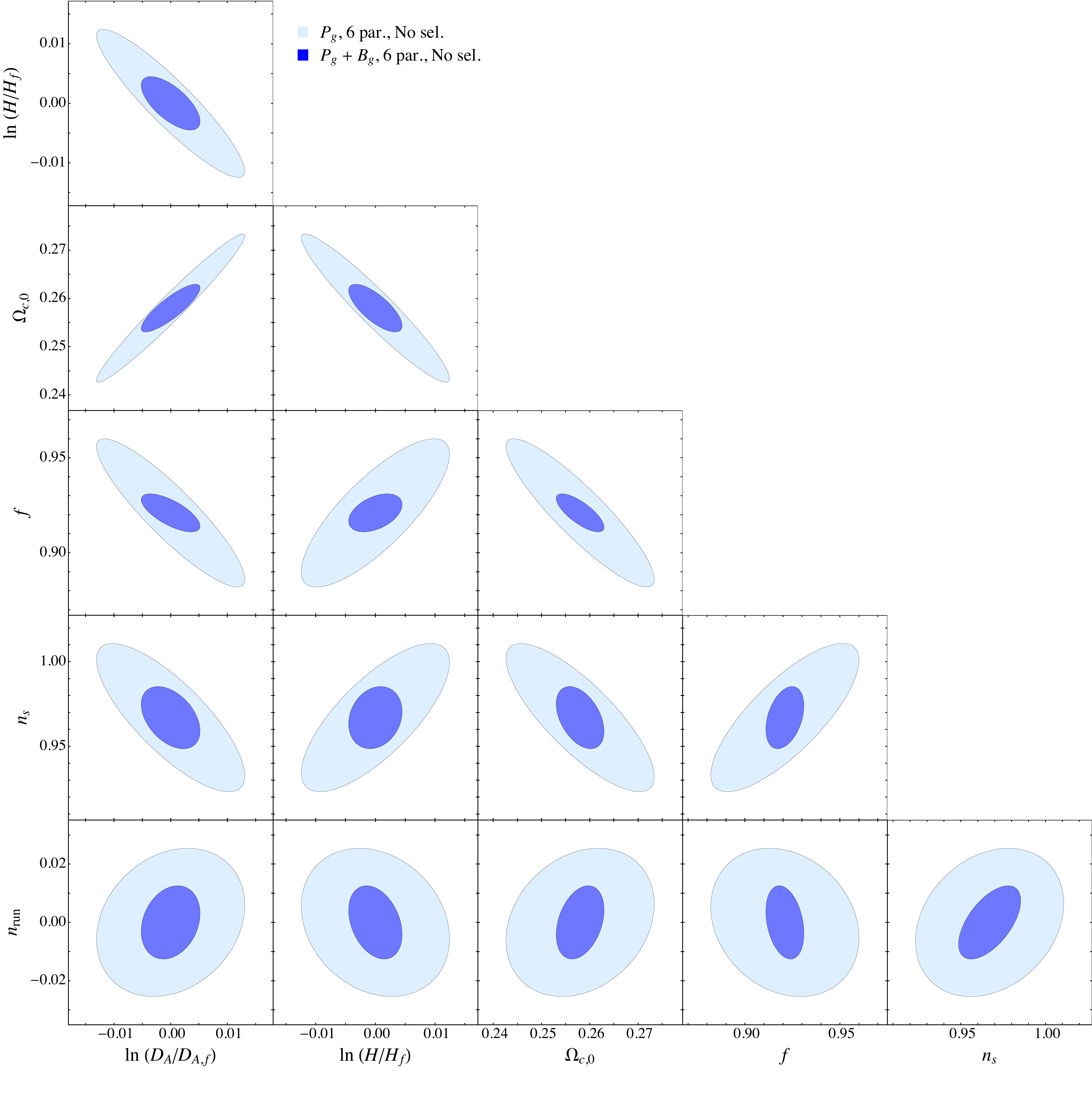}
	\caption{2D contour plots of the projected $1\sigma$ errors for a Euclid-like survey when selection effects are ignored. The bigger light blue ellipses use the power spectrum alone while the smaller dark blue ellipses use both the power spectrum and bispectrum.}
\label{fig:triangleeuclid2}
\end{center}
\end{figure}

We now turn to constraints on all six cosmological parameters and ignoring selection effects. The fiducial values of the model parameters are given in tables\ \ref{table:euclid2} and \ref{table:euclid3}, and the resulting $1\sigma$ errors using the power spectrum alone and including the bispectrum are given in table\ \ref{table:euclid2}. With the power spectrum alone, even in the absence of selection effects, we find that $f$ is strongly degenerate with a few other parameters; in particular, $r = -0.89$ with $\ln D_A$, $-0.90$ with $\Omega_{c0}$, $0.95$ with $P_{\epsilon}^{\{0\}}$, $0.93$ with $b_{K^2}$, and $-0.93$ with $b_{\rm td}$. Including the bispectrum breaks all degeneracies of $f$ and the resulting error estimates on various parameters improve by roughly a factor of 4. Fig.\ \ref{fig:triangleeuclid2} displays the 2D contour plots for the six cosmological parameters in these two cases.

Lastly, we include selection effects in both the power spectrum and bispectrum. In both cases, using the power spectrum alone or including the bispectrum, we again find that the parameters $b_{\rm td}$, $b_{\delta \Pi_{\parallel}^{[2]}}$, $b_{(\Pi^{[2]} K)_{\parallel}}$, and $b_{\Pi_{\parallel}^{[3]}}$ are highly degenerate with one another, with correlations of $r = \pm 1$, and therefore drop $b_{\Pi_{\parallel}^{[3]}}$ from the analysis. The resulting $1\sigma$ errors can be found in table\ \ref{table:euclid3}. In the presence of selection effects, $f$ is strongly-correlated with various bias parameters. Using solely the power spectrum, we again find $r \approx 1.0$ with $b_{\eta}$, while $r = 0.95$ with $b_{K^2}$, $-0.94$ with $b_{\rm td}$, $-0.96$ with $b_{(KK)_{\parallel}}$, and $0.95$ with $b_{(\Pi^{[2]} K)_{\parallel}}$. Upon including the bispectrum most of these degeneracies are broken except that with $b_{\eta}$, with which $f$ has a correlation coefficient $r = 0.99$.

\begin{figure}[!t]
\begin{center}
	\includegraphics[width=6.0in,angle=0]{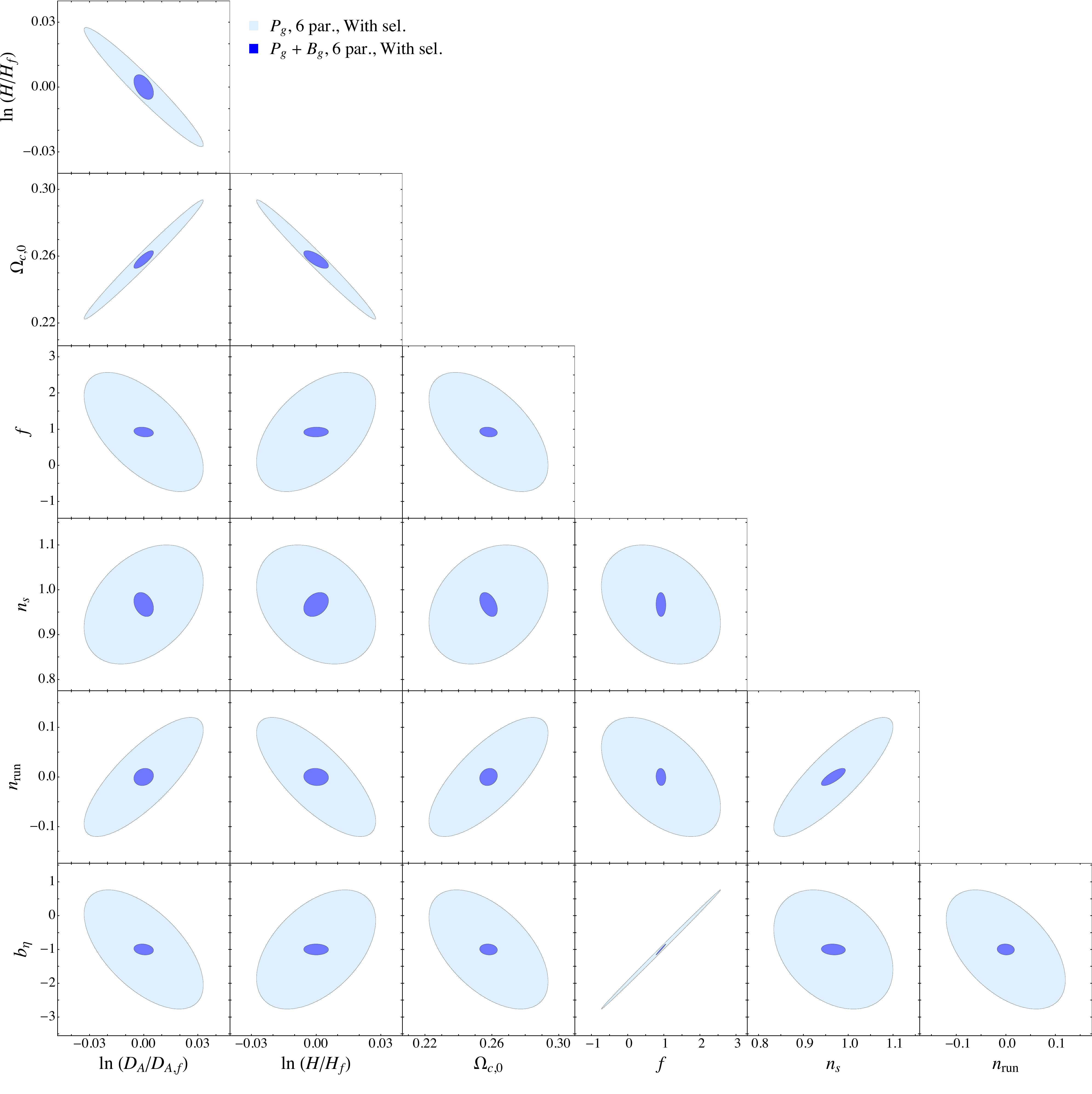}
	\caption{2D contour plots of the projected $1\sigma$ errors for a Euclid-like survey when selection effects are included. The bigger light blue ellipses use the power spectrum alone while the smaller dark blue ellipses use both the power spectrum and bispectrum.}
\label{fig:triangleeuclid3}
\end{center}
\end{figure}

In fig.\ \ref{fig:triangleeuclid3}, we show the 2D contour plots for the six cosmological parameters and $b_{\eta}$ in these two cases. While $b_\eta$ and $f$ are still highly degenerate even when including the bispectrum, we find that interesting constraints on $f$ can nevertheless be placed even after marginalizing over $b_\eta$. As discussed above, the second-order displacement terms which have a characteristic shape dependence in the bispectrum allow for individual constraints on RSDs and selection effects, albeit not at the same level as would be possible without selection contributions.

\begin{figure}[!t]
\begin{center}
	\includegraphics[width=6.0in,angle=0]{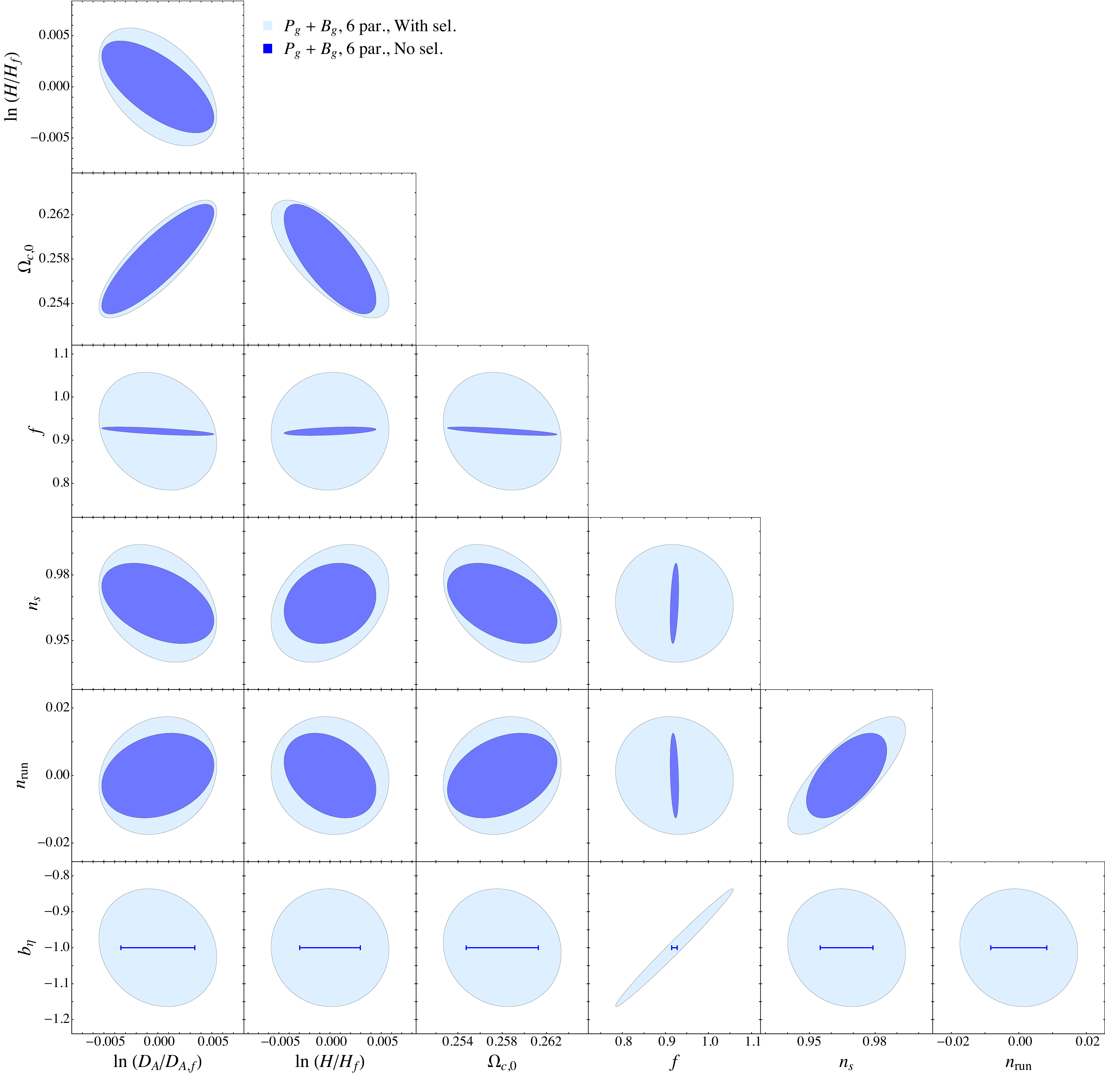}
	\caption{2D contour plots of the projected $1\sigma$ errors for a Euclid-like survey obtained using both the power spectrum and bispectrum. The bigger light blue ellipses include selection effects while the smaller dark blue ellipses ignore them. In the panels that show $b_\eta$, the horizontal $1\sigma$ error bars indicate the case where selection effects are ignored, with $b_{\eta}$ fixed to $-1$.}
\label{fig:triangleeuclid4}
\end{center}
\end{figure}

In fig.\ \ref{fig:triangleeuclid4}, we compare the errors obtained using both the power spectrum and bispectrum when selection effects are ignored (table\ \ref{table:euclid2}, fig.\ \ref{fig:triangleeuclid2}) or included (table\ \ref{table:euclid3}, fig.\ \ref{fig:triangleeuclid3}).

\begin{figure}[!t]
\begin{center}
	\includegraphics[width=6.0in,angle=0]{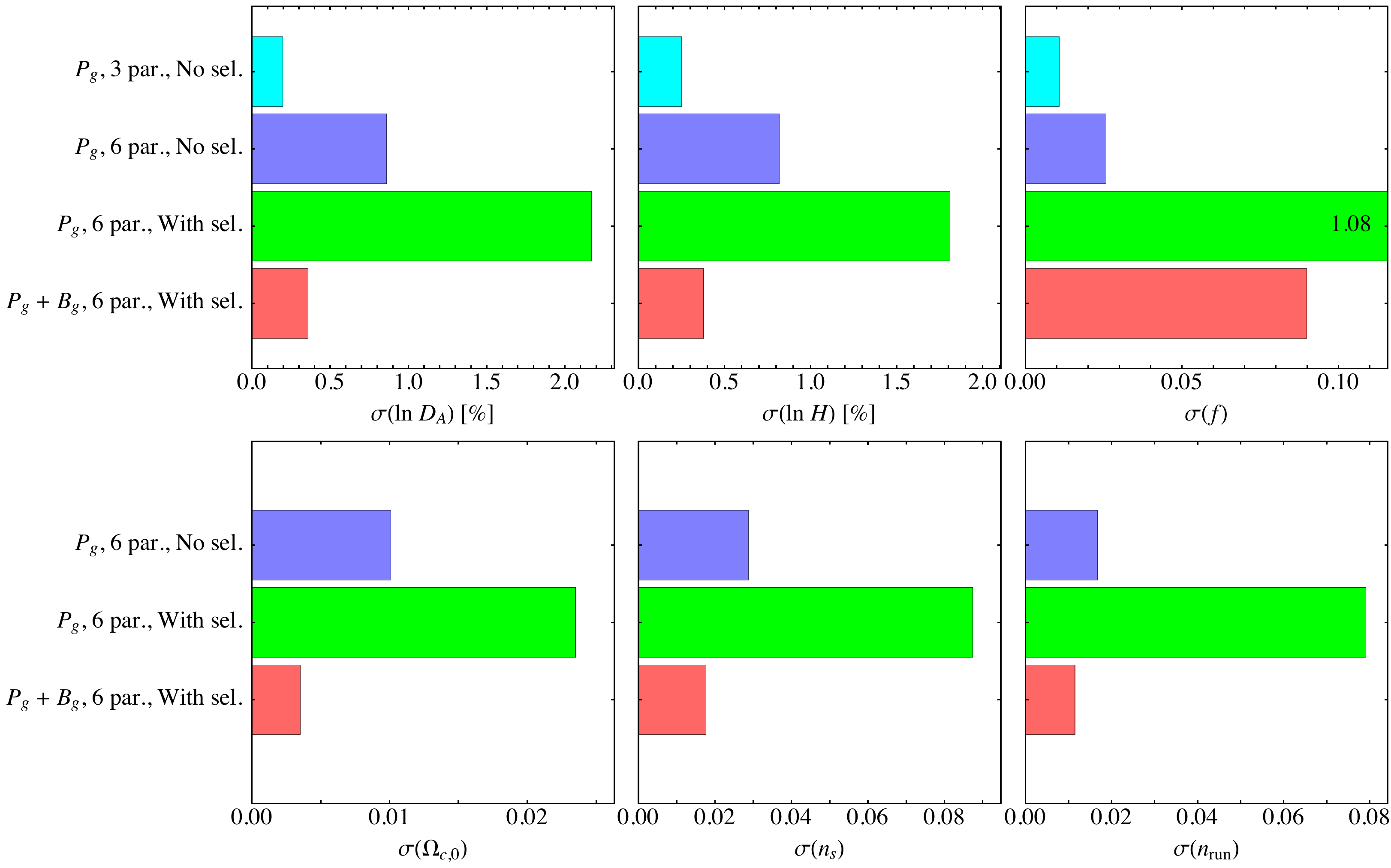}
	\caption{Projected $1\sigma$ errors on the six cosmological parameters in various cases for a Euclid-like survey.}
\label{fig:bareuclid}
\end{center}
\end{figure}

Fig.\ \ref{fig:bareuclid} summarizes the main results of this subsection -- projected errors on the six cosmological parameters -- in the different cases studied above.

\subsubsection{Parameter shifts due to selection effects}

\begin{figure}[!t]
\begin{center}
\includegraphics[width=6.0in,angle=0]{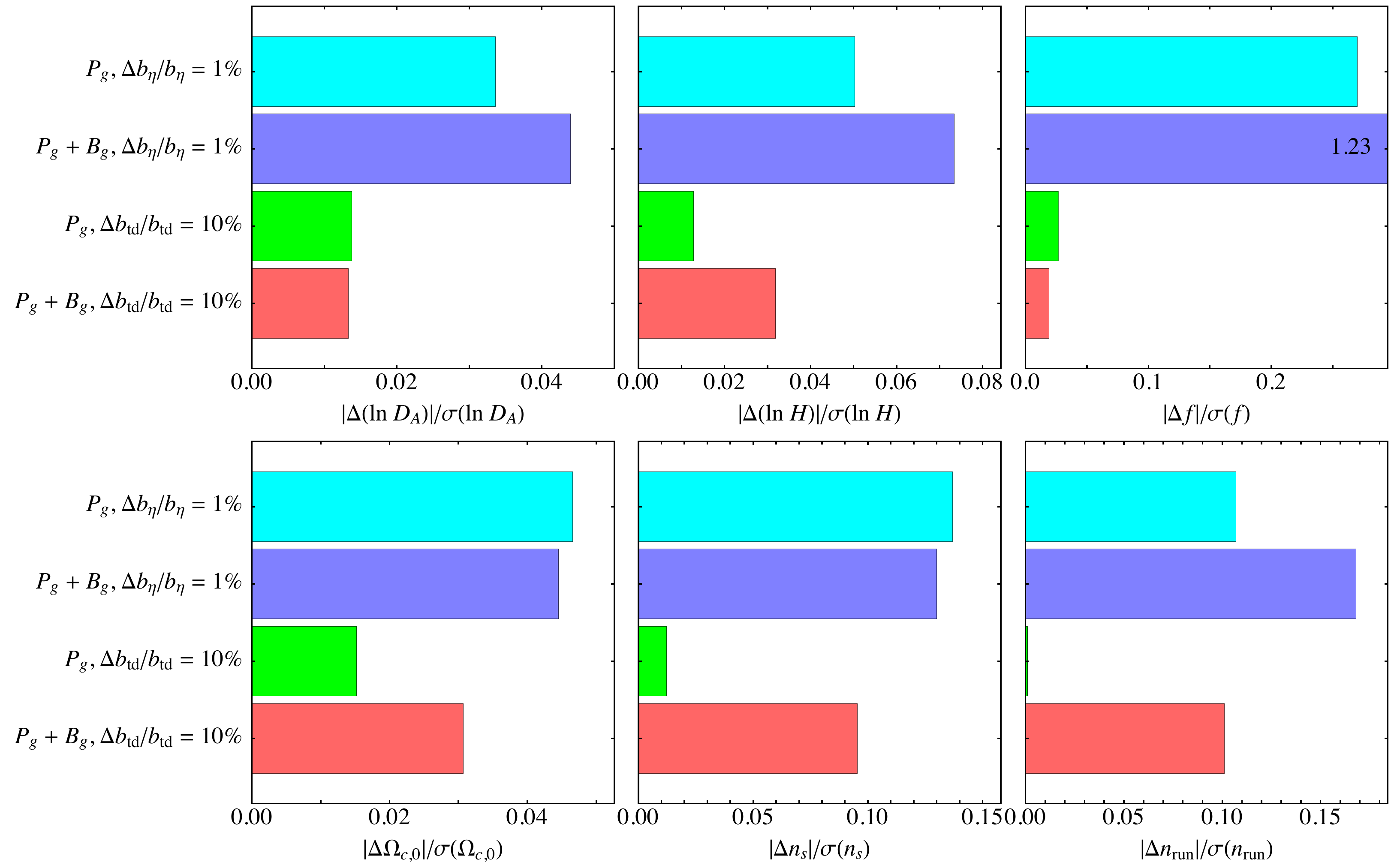}
\caption{Systematic shifts in cosmological parameters, compared to the corresponding $1\sigma$ errors, for a Euclid-like survey when (upper two bars in each panel) $b_\eta$ is fixed to a value that differs from its true value by $1\%$, and no selection effects are included otherwise; and (lower two bars) $b_{\rm td}$ is fixed to a value that differs from its true value by $10\%$ (and no selection effects are included). In each case, we show the results expected for power spectrum only and the combination of power spectrum and bispectrum.
}
\label{fig:bareuclidshifts}
\end{center}
\end{figure}

We have seen that marginalizing over selection effects can significantly degrade cosmological constraints from galaxy clustering, especially if they are solely inferred from the galaxy power spectrum. We will now assess the extent to which cosmological parameters shift if the selection bias parameters are fixed to incorrect values. This can be calculated using eq.\ (\ref{eq:shifts}) and the method described in section\ \ref{sec:shiftsparam}. Table\ \ref{table:euclid4} summarizes the parameter shifts when selection effects are ignored, that is the selection bias parameters are set to the fiducial values given in table\ \ref{table:euclid3} but not marginalized over. For illustration, we assume that $b_{\eta}$ differs from its true value by $1\%$, so that $b_{\eta,{\rm true}} = -0.99$ instead of $-1.0$. We also report the parameter shifts (in case of no selection effects) when $b_{\rm td}$ is fixed to a value that differs from its true value by $10\%$, so that $b_{\rm td,true} = 0.9 \times (23/42)(b_1-1)$ instead of $(23/42)(b_1-1)$. We see that an incorrect estimate of $b_{\eta}$ at the $1\%$ level affects parameter constraints significantly more than a wrong value of $b_{\rm td}$ at the $10\%$ level, the most notable shifts being in $f$. Specifically, we find that a fixed $b_\eta$ differing from the true value by $\sim 4\%$ leads to a $1\sigma$ shift in the estimated growth rate for Euclid when using the power spectrum alone. Upon including the bispectrum, even a $\sim 1\%$ difference in $b_\eta$ leads to a $1\sigma$ shift. All of these results are conveniently summarized in fig.\ \ref{fig:bareuclidshifts}. This demonstrates that selection effects need to either be controlled at the few-percent level or marginalized over.

\subsection{Roman Space Telescope}

For a Roman Space Telescope-like survey, we adopt $z = 1.465$, $V_{\rm survey} = 6.363 \, h^{-3} \, {\rm Gpc}^3$, and $n_g = 2.577 \times 10^{-3} \, h^3 \, {\rm Mpc}^{-3}$, corresponding to the H$\alpha$-emitting galaxies. Compared to Euclid, this survey is centered at the same redshift, albeit with a volume that is 10 times smaller and a number density 5 times larger. For $b_1$ and other bias parameters, we choose the same fiducial values as those adopted for the Euclid-like survey in the previous subsection. For the cosmological parameters, we find the fiducial values $D_A = 1.21 \times 10^3 \, h^{-1} \, {\rm Mpc}$, $H = 230 \, h \, {\rm km \, s^{-1} \, Mpc^{-1}}$, and $f = 0.926$ for the cosmological model considered here. Further, the linear matter power spectrum gives $\sigma_8 = 0.412$.\footnote{The corresponding value of $\sigma_{12}$ is $0.407$.}

We determine $k_{\rm max}$ for the galaxy power spectrum and bispectrum data as before, demanding that theoretical systematics induce shifts no larger than $0.25\sigma$ in all six cosmological parameters. This criterion gives $k_{\rm max} = 0.60 \, h \, {\rm Mpc}^{-1}$ for the power spectrum and $0.13 \, h \, {\rm Mpc}^{-1}$ for the bispectrum. The higher value of $k_{\rm max}$ is due to the generally larger statistical errors on cosmological parameters for Roman, so that larger absolute parameter shifts can be accommodated.

\begin{figure}[!t]
\begin{center}
	\includegraphics[width=6.0in,angle=0]{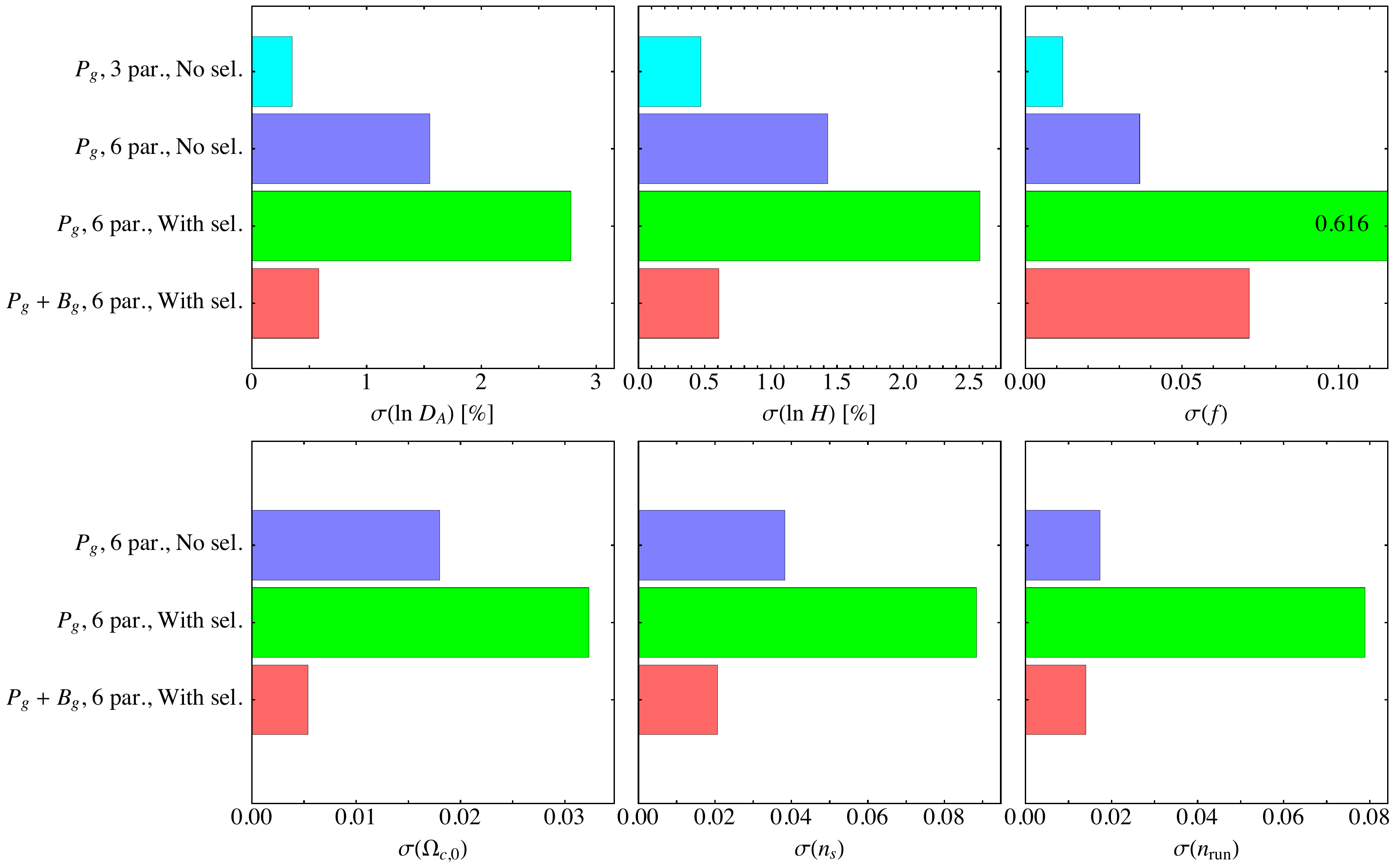}
	\caption{Projected $1\sigma$ errors on the six cosmological parameters in various cases for a Roman Space Telescope-like survey.}
\label{fig:barwfirst}
\end{center}
\end{figure}

We can now obtain the error estimates for various bias and cosmological parameters. We consider first constraints with the power spectrum alone, and varying only three of the six cosmological parameters, $D_A$, $H$, and $f$, keeping $\Omega_{c0}$, $n_s$, and $n_{\rm run}$ fixed. In the absence of selection effects, we find that the growth rate $f$, in particular, is not strongly degenerate with any other parameter. In the presence of selection effects, however, it is again strongly correlated with $b_{\eta}$, with $r = 0.99$. The qualitative behavior of the errors is similar to the outcomes for the Euclid-like survey. Therefore, we do not show similar tables and contour plots for the Roman Space Telescope, but instead summarize the main results for the six cosmological parameters in fig.\ \ref{fig:barwfirst}.

Next, we investigate constraints on all six cosmological parameters in the absence of selection effects. With the power spectrum alone, we find that $f$ is strongly degenerate with a few other parameters; in particular, $r = -0.94$ with $\ln D_A$, $-0.92$ with $\Omega_{c0}$, and $0.92$ with $P_{\epsilon}^{\{0\}}$. Including the bispectrum breaks all degeneracies of $f$ and the resulting error estimates on model parameters improve by roughly a factor of 3. The uncertainties on the six cosmological parameters are shown in fig.\ \ref{fig:barwfirst}.

Lastly, we include selection effects in both the power spectrum and bispectrum. In this case, $f$ is strongly correlated with a number of bias parameters. Using the power spectrum alone we find $r = -0.89$ with $\bI$, $0.89$ with $b_{K^2}$, and $0.99$ with $b_{\eta}$. Upon including the bispectrum, most of these degeneracies are broken except that with $b_{\eta}$, with which $f$ now has $r = 0.96$. The corresponding errors on the six cosmological parameters are also shown in fig.\ \ref{fig:barwfirst}.

\subsection{HETDEX}

For a HETDEX-like survey, we use the survey parameters $z = 2.7$, $V_{\rm survey} = 3.492 \, h^{-3} \, {\rm Gpc}^3$, and $n_g = 3.436 \times 10^{-4} \, h^3 \, {\rm Mpc}^{-3}$, with $b_1 = 2.5$. For the remaining bias parameters, we adopt fiducial values identical to those for Euclid and the Roman Space Telescope. The survey mean redshift implies the fiducial values $D_A = 1.14 \times 10^3 \, h^{-1} \, {\rm Mpc}$, $H = 404 \, h \, {\rm km \, s^{-1} \, Mpc^{-1}}$, and $f = 0.977$ for the cosmological parameters. Furthermore, the linear matter power spectrum gives $\sigma_8 = 0.279$.\footnote{The corresponding value of $\sigma_{12}$ is $0.276$.}

An analysis similar to that carried out in the previous subsections leads to $k_{\rm max} \simeq 0.87 \, h \, {\rm Mpc}^{-1}$ and $0.20 \, h \, {\rm Mpc}^{-1}$ for the galaxy power spectrum and bispectrum respectively. These significantly higher values of $k_{\rm max}$ are expected for HETDEX, since the much higher redshift increases the reach of perturbation theory, and since the larger statistical error bars of HETDEX (due to the smaller volume) allow for larger absolute shifts. The wide range in wavenumber also results in fewer degeneracies between the six cosmological parameters and various nonlinear bias parameters, as we will see shortly.

\begin{figure}[!t]
\begin{center}
	\includegraphics[width=6.0in,angle=0]{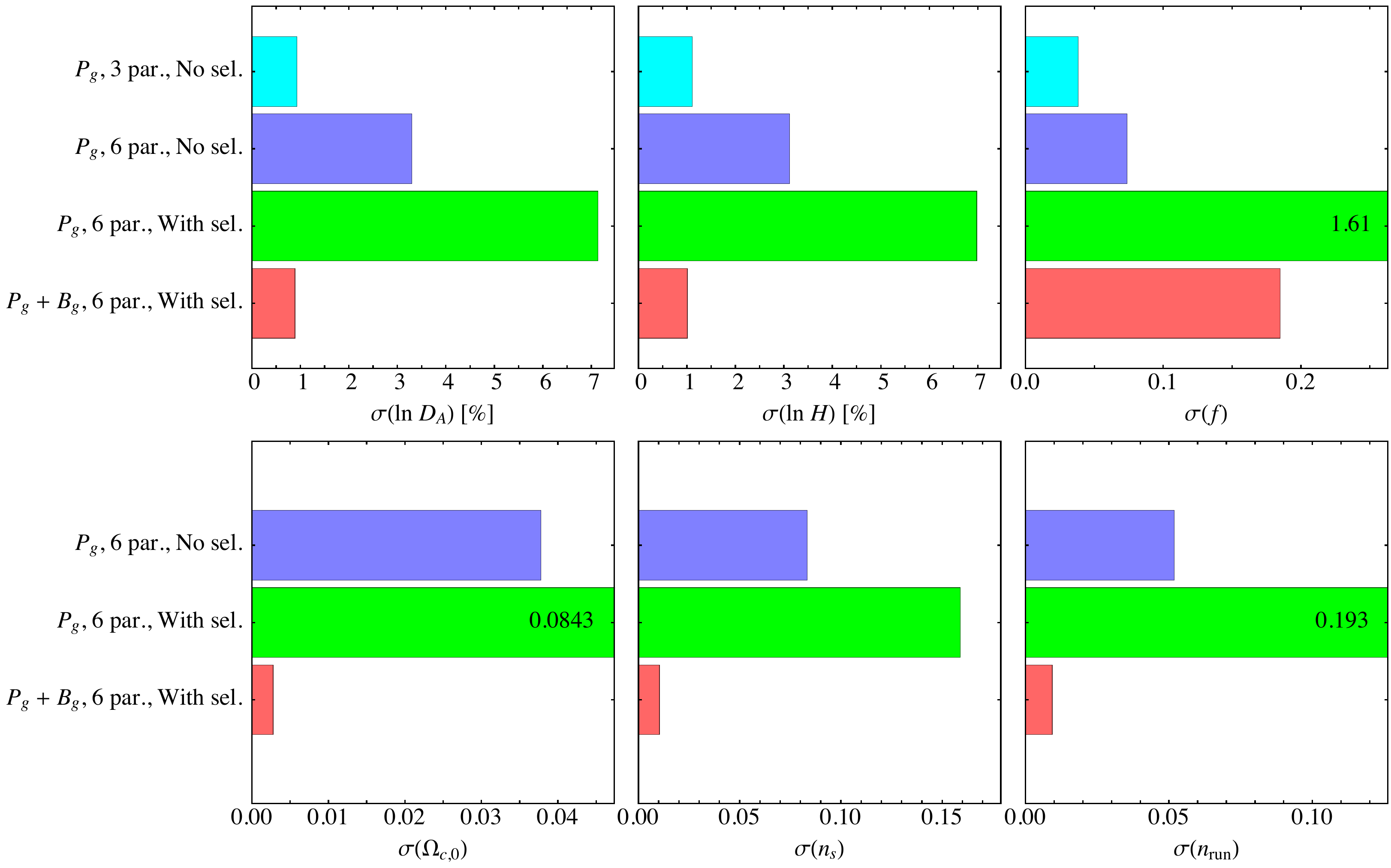}
	\caption{Projected $1\sigma$ errors on the six cosmological parameters in various cases for a HETDEX-like survey.}
\label{fig:barhetdex}
\end{center}
\end{figure}

We begin again with constraints from the power spectrum alone, varying only three of the six cosmological parameters, $D_A$, $H$, and $f$, while $\Omega_{c0}$, $n_s$, and $n_{\rm run}$ are held fixed. In the absence of selection effects, the growth rate $f$, in particular, is not strongly degenerate with any other parameter. In the presence of selection effects however, it does correlate strongly with a couple of parameters; we find $r \approx 1.0$ with $b_{\eta}$, and $0.90$ with $b_{(\Pi^{[2]} K)_{\parallel}}$. The qualitative behavior of the errors is, again, similar to that found for Euclid. We summarize the main results for the six cosmological parameters in fig.\ \ref{fig:barhetdex}.

When all six cosmological parameters are simultaneously fitted for, and ignoring selection effects, we find that using either the power spectrum alone or including the bispectrum, $f$ is not strongly degenerate with any other parameter. Including the bispectrum improves error estimates on the model parameters by roughly a factor of 4. The errors on the six cosmological parameters are shown in fig.\ \ref{fig:barhetdex}.

Finally, when selection effects are included in both the power spectrum and bispectrum, $f$ strongly correlates only with $b_{\eta}$. Using the power spectrum alone yields $r = 0.99$, whereas including the bispectrum gives $r = 0.98$. The corresponding errors on the six cosmological parameters are also shown in fig.\ \ref{fig:barhetdex}.

\section{Conclusions}
\label{sec:conc}

In order to extract the maximum amount of information from current and upcoming galaxy surveys, it is crucial to correctly model the nonlinear growth of structure. While the leading-order power spectrum alone is insufficient to break parameter degeneracies such as that between $f$ and $\sigma_8$, higher-order perturbative contributions and higher-order statistics such as the bispectrum can help in doing so. Modeling the nonlinear regime, however, requires additional parameters beyond the linear bias and shot-noise amplitude, such as higher-order and higher-derivative bias parameters. In this paper we have studied the constraints on cosmological parameters in a comprehensive model of nonlinear bias which includes all these terms. We focused in particular on the line-of-sight dependent selection effects \cite{Desjacques:2018pfv}, which had not been included in forecast studies previously. We assessed the amount of additional cosmological information that can be extracted from the galaxy bispectrum relative to a simple power-spectrum only analysis, as in \citep{Sefusatti:2006pa,Yankelevich:2018uaz,Gualdi:2020ymf}, while attempting to self-consistently exclude scales $k > k_{\rm max}$ that are affected by even higher-order corrections.

\begin{figure}[!t]
\begin{center}
	\includegraphics[width=6.0in,angle=0]{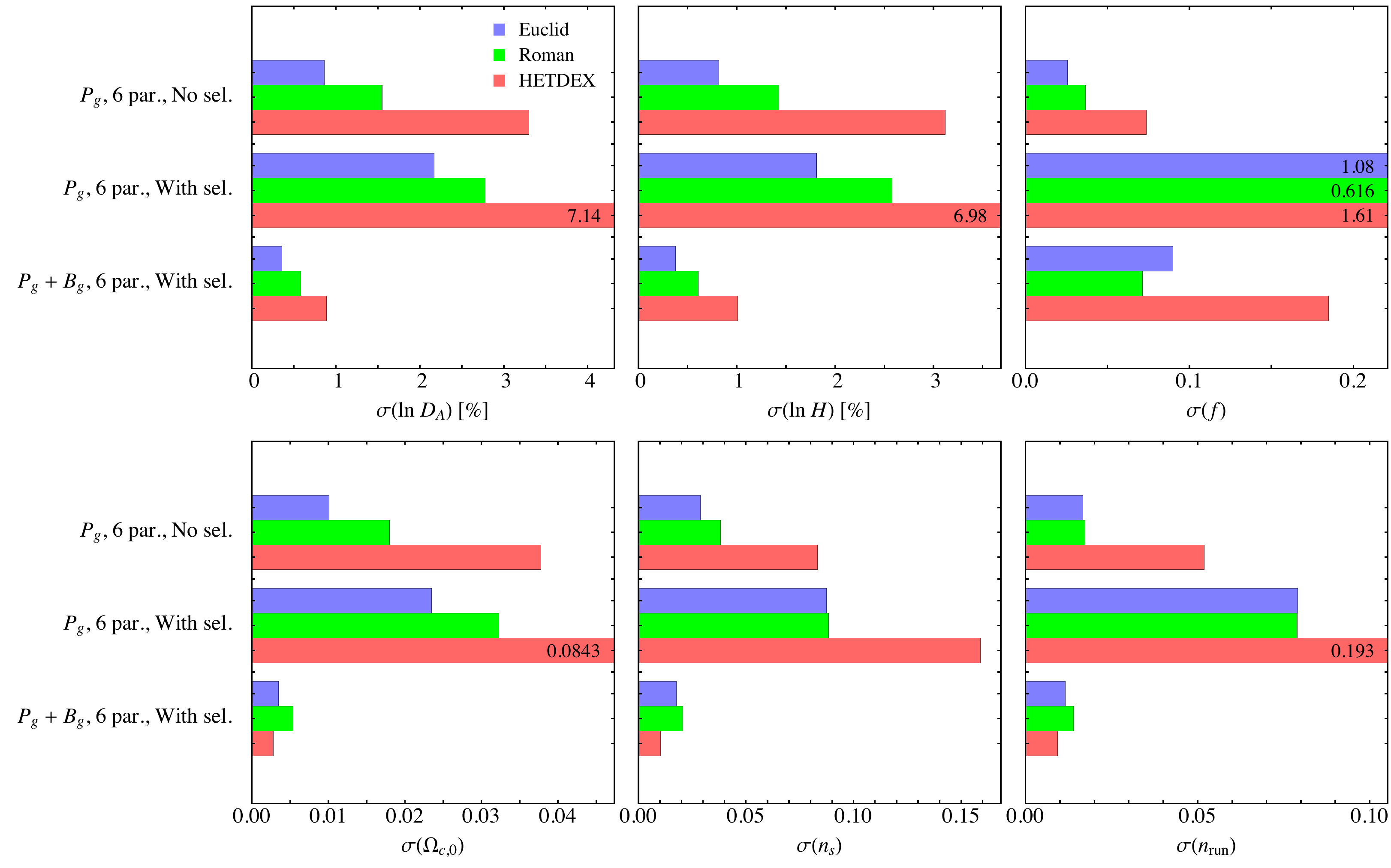}
	\caption{Summary of projected $1\sigma$ errors on the six cosmological parameters in various cases for all three surveys.}
\label{fig:barall}
\end{center}
\end{figure}

In particular, we obtained Fisher constraints on the six cosmological parameters $\{D_A, H,$ $\Omega_{c0}, f, n_s, n_{\rm run}\}$ using the 1-loop galaxy power spectrum and the tree-level galaxy bispectrum. We introduced a self-consistent method to determine the maximum wavenumber $k_{\rm max}$ to be used in error forecasts, from the requirement that parameter shifts that arise from ignoring the next-order correction in perturbation theory are less than a given fraction of the statistical errors. In fig.\ \ref{fig:barall}, we summarize our results in the different cases considered, using the power spectrum alone or including the bispectrum and with or without selection effects, for three surveys: Euclid, the Roman Space Telescope, and HETDEX.

Our three survey configurations qualitatively yield similar results. Let us summarize the main takeaway points for a Euclid-like survey:
\begin{itemize}
	\item In the classic, three-parameter power spectrum-only analysis, allowing for selection effects increases the error in $f$ (equivalently $f\sigma_8$) by about a factor of 80 due to the perfect degeneracy between the selection bias $b_\eta$ and the linear-order RSD contribution. The errors on $\ln D_A$ and $\ln H$, on the other hand, increase by $25-35\%$.
	\item Upon expanding the cosmological parameters space from three to six, the error on $f$ increases by a factor of two in the absence of selection effects. Further including selection effects results in the loss of any constraining power on $f$ as the error becomes of the order of $f$ itself.
	\item Including the bispectrum breaks the degeneracy between $f$ and $\sigma_8$, thanks to second-order displacement terms. In the absence of selection effects, the error on $f$ shrinks by roughly a factor of 4.
	\item The bispectrum also breaks the degeneracy between $f$ and selection effects. If one allows for all selection bias terms to be free, then the constraint on $f$ is at the $10\%$ level, a factor of 14 times worse than the combined power spectrum and bispectrum analysis in the absence of selection effects, albeit only around 3 times worse than the power spectrum-only analysis without selection effects.
	\item Fixing selection bias parameters to incorrect values can lead to biased constraints on cosmological parameters. In particular, if $b_\eta$ is fixed, then a few-percent systematic error in this parameter can cause a $1\sigma$ shift in the estimated value of $f$. 
\end{itemize}

Our precise numerical results should be taken with a bit of skepticism as is usual with Fisher forecasts. Moreover, while our determination of $k_{\rm max}$ is in principle self-consistent, it is only a rough approximation as we do not have prior knowledge on the relevant higher-order bias parameters. There are also a number of caveats regarding our Gaussian and diagonal covariance approximation (albeit using a nonlinear power spectrum) with white noise. First, we have ignored correlations among modes that arise from sparse sampling or, simply, the survey window function. These can lead to the breaking of the Gaussian likelihood approximation \cite{Scoccimarro:2000sn}. Second, there are non-Gaussian contributions to the power spectrum covariance (which involves the trispectrum) and bispectrum covariance (which involves connected $n$-point functions up to the 6-point function). These will also generate off-diagonal contributions to the covariance. For the survey configurations considered here, however, recent studies \cite{Wadekar:2019rdu,Blot:2018oxk} suggest that the covariance of power spectrum multipoles on weakly nonlinear scales is dominated by shot-noise and super-survey mode coupling rather than non-Gaussian terms. Third, one should also take into account the cross-covariance between the power spectrum and the bispectrum. Note, however, that this effect was found to be small in the Fisher matrix analysis of \cite{Yankelevich:2018uaz}.

Our forecast for the constraining power of the galaxy bispectrum in the case of a Euclid-like survey is significantly more optimistic than that of \cite{Yankelevich:2018uaz} despite the fact that we adopted a similar $k_\text{max}$ for the bispectrum. In particular, we find that combining the two statistics ($P_g + B_g$) substantially improves constraints on the cosmological parameters even when selection effects are included. The reason for this discrepancy presumably lies in the fact that \cite{Yankelevich:2018uaz} split the surveyed volume into 14 redshift bins, each with an independent set of bias parameters, leading to a total of 56 bias parameters (which do not even fully characterize the galaxy power spectrum at one-loop) to be marginalized over. In our opinion, this approach is likely too conservative since it is physically expected that the bias parameters evolve smoothly across the surveyed redshift range. It may be more appropriate, therefore, to consider a parametric approach that takes into account the expected continuity of the redshift dependence of the galaxy bias parameters. On the other hand, we have only considered a single set of bias parameters here. The realistic case will thus be a compromise between these two extremes. Let us also stress that compression methods, such as those proposed in, for example, \cite{Vogeley:1996xu,Byun:2017fkz,Gualdi:2017iey}, could be applied to the data in order to optimize the extraction of cosmological information.

It is also worth noting that while it is important to include the 1-loop power spectrum in order to break the degeneracy between $f$ and $b_{\eta}$ that is present at linear order, the various 1-loop terms are likely not all independent. This is suggested by the fact that we found the parameters $b_{\rm td}$, $b_{\delta \Pi_{\parallel}^{[2]}}$, $b_{(\Pi^{[2]} K)_{\parallel}}$, and $b_{\Pi_{\parallel}^{[3]}}$ to be highly degenerate with one another (we dropped $b_{\Pi_{||}}^{[3]}$ from our analysis for this reason). One should therefore carry out a principal component analysis to find which terms (and bias parameters), or combinations thereof, are distinguishable, and only include those in the analysis. Doing so should help reduce the parameter space and improve sampling efficiency in real analyses.

Our results show that a careful control of selection biases (which were not taken into account in the Fisher forecasts of \cite{Sefusatti:2006pa,Yankelevich:2018uaz,Gualdi:2020ymf}) is crucial. Ignoring selection effects is not an option for Stage-IV galaxy redshift surveys and some form of marginalization over selection bias parameters is likely necessary. This will, in turn, directly impact constraints on $f$ or $f\sigma_8$. Including the bispectrum, however, helps in recovering a significant fraction of the cosmological information by breaking parameter degeneracies, in particular thanks to the second-order displacement terms that are protected by the equivalence principle.

The Fisher code used for the results in this paper is available at \href{https://bitbucket.org/nishant_agarwal/fisher_lss_p-b/src/master/}{this URL}.


\acknowledgments 
N.~A. thanks Aoife Boyle for useful discussions. V.~D. acknowledges support from the Israel Science Foundation (grant no. 1395/16). D.~J. acknowledges support from the National Science Foundation grant AST-1517363 and NASA ATP program (80NSSC18K1103). F.~S. acknowledges support from the Starting Grant (ERC-2015-STG 678652) ``GrInflaGal'' of the European Research Council. 


\appendix
\section{Derivatives $\partial \ln P_g/\partial \theta^a$ in eq.\ (\ref{eq:pkfisher})}
\label{app:pkderivatives}

\renewcommand{\theequation}{A\arabic{equation}}
\setcounter{equation}{0}

In this appendix we consider the derivatives that appear in the Fisher analysis, eq.\ (\ref{eq:pkfisher}). We compute logarithmic derivatives of the galaxy power spectrum $P_g(k,\mu)$ in eq.\ (\ref{eq:pgsfull}) with respect to the six cosmological parameters $\{D_A, H, \Omega_{c0}, f, n_s, n_{\rm run}\}$; those with the remaining parameters that describe galaxy clustering are straightforward to calculate from eqs.\ (\ref{eq:pgsfull}) to (\ref{eq:pgs13}). The derivatives with $\ln D_A$ and $\ln H$ are calculated in a two-step process and using $k^2 = \( \frac{H_f}{H} \)^2 k_{{\rm true},\parallel}^2 + \( \frac{D_A}{D_{A,f}} \)^2 k_{{\rm true},\perp}^2$, where $k$ is the observed wavenumber for fiducial values $D_{A,f}$ and $H_f$, $k_{\rm true}$ is the true wavenumber for the parameters $D_A$ and $H$, and parallel and perpendicular are defined according to the line-of-sight direction \cite{Shoji:2008xn}. All derivatives are calculated with respect to the cosmological parameters and at the point that they equal the fiducial values. We find that
\bea
	\frac{\partial \ln P_g(k,\mu)}{\partial \ln D_A} & = & \frac{\partial \ln P_g(k,\mu)}{\partial \ln k} \frac{\partial \ln k}{\partial \ln D_A} + \frac{\partial \ln P_g(k,\mu)}{\partial \mu^2} \frac{\partial \mu^2}{\partial \ln D_A} \, ,
\eea
\bea
	\frac{\partial \ln P_g(k,\mu)}{\partial \ln H} & = & \frac{\partial \ln P_g(k,\mu)}{\partial \ln k} \frac{\partial \ln k}{\partial \ln H} + \frac{\partial \ln P_g(k,\mu)}{\partial \mu^2} \frac{\partial \mu^2}{\partial \ln H} \, ,
\eea
with
\bea
	\frac{\partial \ln k}{\partial \ln D_A} & = & 1 - \mu^2 \, , \\
	\frac{\partial \mu^2}{\partial \ln D_A} & = & -2\mu^2 (1 - \mu^2) \, , \\
	\frac{\partial \ln k}{\partial \ln H} & = & - \mu^2 \, , \\
	\frac{\partial \mu^2}{\partial \ln H} & = & -2\mu^2 (1 - \mu^2) \, .
\eea
We also need
\bea
	\frac{\partial \ln P_g(k,\mu)}{\partial \ln k} & = & \, n_{\rm eff}(k,\mu) \, ,
\eea
\bea
	& & \frac{\partial \ln P_g(k,\mu)}{\partial \mu^2} \, = \, \frac{1}{P_g(k,\mu)} \Bigg[ -2 b_{\eta}f \( b_1 - b_{\eta} \mu^2 f \) P_L(k) + 2 \Big\{ b_{\eta} f ( b_{\nabla^2\delta} + b_1 \bI \nonumber \\
	& & \qquad \quad + \ 2b_1 \bII \mu^2 ) - b_{\eta}^2 \mu^2 f^2 \( 2\bI + 3\bII \mu^2 \) \Big\} k^2 P_L(k) + b_{\eta} k^2 P_{\eps\varepsilon_{\eta}}^{\{2\}} \nonumber \\
	& & \qquad + \, \sum_{n=0}^4 \sum_{(m,p)} A_{nmp} (f,\{b_{O}\}_{2-2}) {\cal I}_{mp}(k) n \mu^{2(n-1)} \nonumber \\
	& & \qquad + \, 2\left\{ \sum_{l=0}^3 \sum_{n=1}^5 C_n^{1-3,\ell} (f,\{b_{O}\}_{2-2}) {\cal I}_{n}(k) \frac{\d {\cal L}_{2\ell}(\mu)}{\d\mu^2} \right\} P_L(k) \Bigg] \, ,
\eea
where we have defined an effective spectral index $n_{\rm eff}(k,\mu)$,
\bea
	& & n_{\rm eff}(k,\mu) \, = \, \frac{1}{P_g(k,\mu)} \Bigg[ \( b_1 - b_{\eta} \mu^2 f \)^2 P_L(k) n_L(k) \nonumber \\
	& & \qquad - \ 2 \Big\{ b_1 b_{\nabla^2\delta} - b_{\eta} \mu^2 f \( b_{\nabla^2\delta} + b_1 \bI + b_1 \bII \mu^2 \) \nonumber \\
	& & \qquad \quad + \ b_{\eta}^2 \mu^4 f^2 \( \bI + \bII \mu^2 \) \Big\} k^2 P_L(k) \{ 2 + n_L(k) \} + 2k^2 P_{\eps}^{\{2\}} + 2b_{\eta} \mu^2 k^2 P_{\eps\varepsilon_{\eta}}^{\{2\}} \nonumber \\
	& & \qquad + \, \sum_{n=0}^4 \sum_{(m,p)} A_{nmp} (f,\{b_{O}\}_{2-2}) {\cal I}_{mp}(k) n_{mp}(k) \mu^{2n} \nonumber \\
	& & \qquad + \, 2\left\{ \sum_{l=0}^3 \sum_{n=1}^5 C_n^{1-3,\ell} (f,\{b_{O}\}_{1-3}) {\cal I}_{n}(k) {\cal L}_{2\ell}(\mu) \right\} P_L(k) n_L(k) \nonumber \\
	& & \qquad + \, 2\left\{ \sum_{l=0}^3 \sum_{n=1}^5 C_n^{1-3,\ell} (f,\{b_{O}\}_{1-3}) {\cal I}_{n}(k) n_n(k) {\cal L}_{2\ell}(\mu) \right\} P_L(k) \Bigg] \, ,
\eea
with $n_L(k) = \frac{\partial \ln P_L(k)}{\partial \ln k}$, $n_{mp}(k) = \frac{\partial \ln {\cal I}_{mp}(k)}{\partial \ln k}$, and $n_n(k) = \frac{\partial \ln {\cal I}_n(k)}{\partial \ln k}$. We calculate the derivative with $\Omega_{c0}$ numerically, using
\bea
	& & \frac{\partial \ln P_g(k,\mu)}{\partial \Omega_{c0}} \, = \, \frac{1}{P_g(k,\mu)} \frac{\Delta P_g(k,\mu)}{\Delta \Omega_{c0}} \, ,
\label{eq:deroc0pk}
\eea
where $\Delta P_g(k,\mu)$ is the difference of $P_g(k,\mu)$ for values of $\Omega_{c0}$ slightly above and below the fiducial value. We modify $\Omega_{\Lambda 0}$ accordingly as we change $\Omega_{c0}$ to keep the Universe flat. Note that $\Omega_{c0}$ enters through the growth factor, growth rate, linear matter power spectrum $P_L(k)$, and loop integrals ${\cal I}_{mp}(k)$ and ${\cal I}_{n}(k)$. The next derivative, with $f$, is given by
\bea
	& & \frac{\partial \ln P_g(k,\mu)}{\partial f} \, = \, \frac{1}{P_g(k,\mu)} \Bigg[ -2b_{\eta} \mu^2 \( b_1 - b_{\eta} \mu^2 f \) P_L(k) \nonumber \\
	& & \qquad + \ 2 \Big\{ b_{\eta} \mu^2 \( b_{\nabla^2\delta} + b_1 \bI + b_1 \bII \mu^2 \) + 2 b_{\eta}^2 \mu^4 f \( \bI + \bII \mu^2 \) \Big\} k^2 P_L(k) \nonumber \\
	& & \qquad + \, \sum_{n=0}^4 \sum_{(m,p)} \frac{\partial A_{nmp} (f,\{b_{O}\}_{2-2})}{\partial f} {\cal I}_{mp}(k) \mu^{2n} \nonumber \\
	& & \qquad + \, 2\left\{ \sum_{l=0}^3 \sum_{n=1}^5 \frac{\partial C_n^{1-3,\ell} (f,\{b_{O}\}_{1-3})}{\partial f} {\cal I}_{n}(k) {\cal L}_{2\ell}(\mu) \right\} P_L(k) \Bigg] \, ,
\label{eq:derf}
\eea
and lastly the derivatives with $n_s$ and $n_{\rm run}$ are also calculated numerically, analogous to the derivative with $\Omega_{c0}$ in eq.\ (\ref{eq:deroc0pk}). Note that $n_s$ and $n_{\rm run}$ only enter through the linear matter power spectrum $P_L(k)$ and loop integrals ${\cal I}_{mp}(k)$ and ${\cal I}_{n}(k)$.

\begin{figure}[!t]
\begin{center}
	\includegraphics[width=2.85in,angle=0]{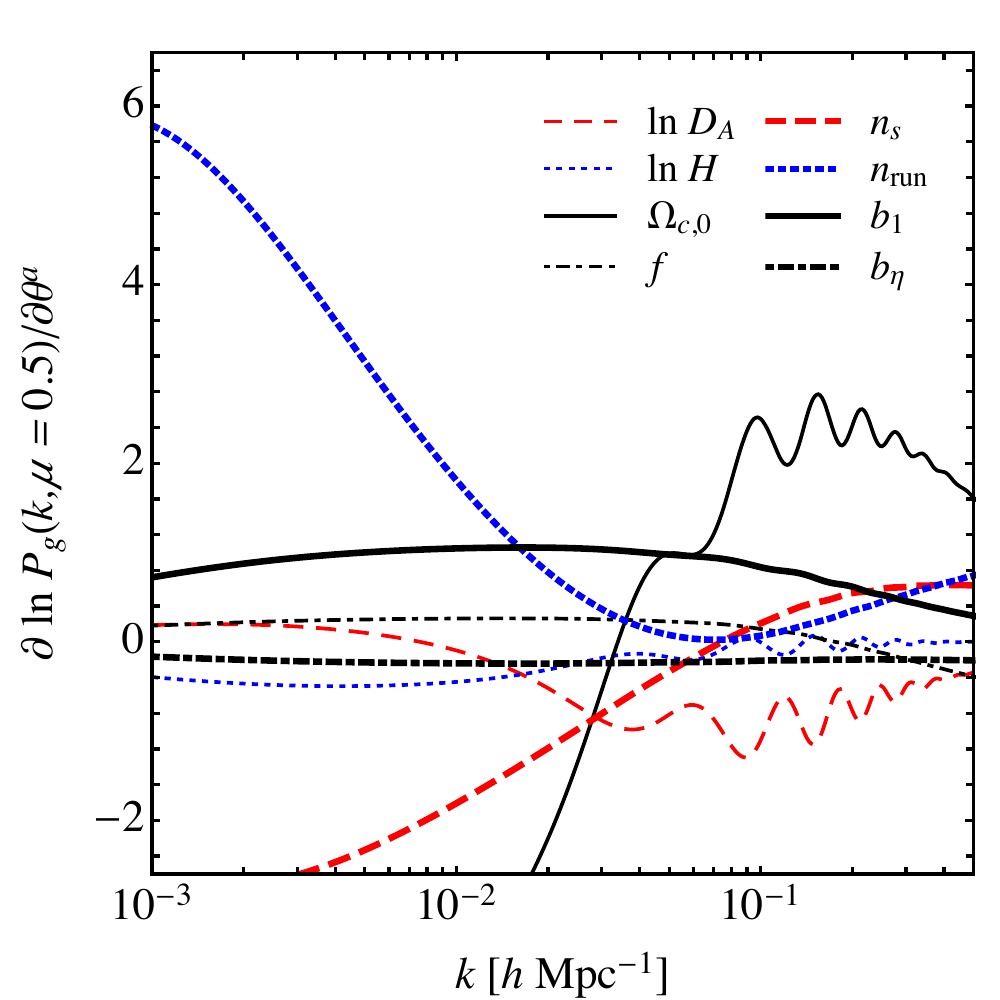}
	\caption{Derivatives of the galaxy power spectrum with $\mu = 0.5$ for a Euclid-like survey.}
\label{fig:pkdereuclid}
\end{center}
\end{figure}

As an example, we show how the derivatives of the power spectrum with respect to the six cosmological parameters and $b_1$ and $b_{\eta}$ vary with $k$ for an arbitrary value of $\mu = 0.5$ for a Euclid-like survey in fig.\ \ref{fig:pkdereuclid}. The small-scale power is expected to break various parameter degeneracies.

\section{Derivatives $\partial B_g/\partial \theta^a$ in eq.\ (\ref{eq:bkfisher})}
\label{app:bkderivatives}

\renewcommand{\theequation}{B\arabic{equation}}
\setcounter{equation}{0}

In this appendix we consider the derivatives that appear in the Fisher analysis, eq.\ (\ref{eq:bkfisher}). We compute derivatives of the galaxy bispectrum $B_g(\vk_1,\vk_2,\vk_3)$ in eq.\ (\ref{eq:bgs}) with respect to the six cosmological parameters $\{D_A, H, \Omega_{c0}, f, n_s, n_{\rm run}\}$; those with the remaining parameters that describe galaxy clustering are straightforward to calculate from eq.\ (\ref{eq:bgs}). The derivatives with $\ln D_A$ and $\ln H$ are calculated in a two-step process as we did for the galaxy power spectrum in appendix \ref{app:pkderivatives} \cite{Greig:2012zw}. We find that
\bea
	\frac{\partial B_g(\vk_1,\vk_2,\vk_3)}{\partial \ln D_A} & = & \sum_{i=1}^3 \( k_i \frac{\partial B_g}{\partial k_i} \frac{\partial \ln k_i}{\partial \ln D_A} + \frac{\partial B_g}{\partial \mu_i} \frac{\partial \mu_i}{\partial \ln D_A} \) ,
\eea
\bea
	\frac{\partial B_g(\vk_1,\vk_2,\vk_3)}{\partial \ln H} & = & \sum_{i=1}^3 \( k_i \frac{\partial B_g}{\partial k_i} \frac{\partial \ln k_i}{\partial \ln H} + \frac{\partial B_g}{\partial \mu_i} \frac{\partial \mu_i}{\partial \ln H} \) ,
\eea
with
\bea
	\frac{\partial \ln k_i}{\partial \ln D_A} & = & 1 - \mu_i^2 \, , \\
	\frac{\partial \mu_i}{\partial \ln D_A} & = & -\mu_i (1 - \mu_i^2) \, , \\
	\frac{\partial \ln k_i}{\partial \ln H} & = & - \mu_i^2 \, , \\
	\frac{\partial \mu_i}{\partial \ln H} & = & -\mu_i (1 - \mu_i^2) \, .
\eea
Note that we take partial derivatives with all six parameters, $k_1$, $k_2$, $k_3$, $\mu_1$, $\mu_2$, and $\mu_3$, one at a time while keeping the other five fixed, even though they are not all independent. Before taking the derivatives we need to replace, for example, $\mu_{23}$ inside $S_{O} (\vk_2,\vk_3,f)$ in eq.\ (\ref{eq:bgs}) with $\( k_1^2 - k_2^2 - k_3^2 \)/(2k_2 k_3)$. The derivatives $\partial B_g/\partial k_i$ and $\partial B_g/\partial \mu_i$ are then straightforward to calculate from eq.\ (\ref{eq:bgs}), where we also set $\partial P_L(k_i)/\partial k_i = P_L(k_i) n_L(k_i)$. We calculate the derivative with $\Omega_{c0}$ numerically, using
\bea
	& & \frac{\partial B_g(\vk_1,\vk_2,\vk_3)}{\partial \Omega_{c0}} \, = \, \frac{\Delta B_g(\vk_1,\vk_2,\vk_3)}{\Delta \Omega_{c0}} \, ,
\label{eq:deroc0bg}
\eea
where $\Delta B_g(\vk_1,\vk_2,\vk_3)$ is the difference of $B_g(\vk_1,\vk_2,\vk_3)$ for values of $\Omega_{c0}$ slightly above and below the fiducial value. We modify $\Omega_{\Lambda 0}$ accordingly as we change $\Omega_{c0}$ to keep the Universe flat. Note that $\Omega_{c0}$ enters through the growth factor, growth rate, and linear matter power spectrum $P_L(k)$. The next derivative, with $f$, is given by
\bea
	& & \frac{\partial B_g(\vk_1,\vk_2,\vk_3)}{\partial f} = -2b_{\eta} \[ \mu_2^2 \(b_1 - b_{\eta} \mu_3^2 f\) + \mu_3^2 \(b_1 - b_{\eta} \mu_2^2 f\) \] \sum_{\{b_O\}_{2-2}} b_{O} S_{O} (\vk_2,\vk_3,f) \nonumber \\
	& & \qquad \times \ P_L(k_2) P_L(k_3) + 2\(b_1 - b_{\eta} \mu_2^2 f\) \(b_1 - b_{\eta} \mu_3^2 f\) \sum_{\{b_O\}_{2-2}} b_{O} \frac{\partial S_{O} (\vk_2,\vk_3,f)}{\partial f} P_L(k_2) P_L(k_3) \nonumber \\
	& & \qquad - \ 2b_{\eta} \mu_1^2 P_{\epsilon\epsilon_{\delta}}^{\{0\}} P_L(k_1) - 2 \(b_1 - 2 b_{\eta} \mu_1^2 f\) \mu_1^2 P_{\epsilon\epsilon_{\eta}}^{\{0\}} P_L(k_1) \, + \, {\rm 2 \ perm.} \, ,
\eea
and lastly the derivatives with $n_s$ and $n_{\rm run}$ are also calculated numerically, analogous to the derivative with $\Omega_{c0}$ in eq.\ (\ref{eq:deroc0bg}). Note that $n_s$ and $n_{\rm run}$ only enter through the linear matter power spectrum $P_L(k)$.


\newpage

\section{Forecast details: Euclid}
\label{app:tableresults}

\renewcommand{\thetable}{C\arabic{table}}
\setcounter{table}{0}

\begin{table}[!h]
  \centering
\begin{center}
	\begin{tabular}{|c|c|c|c|}
	\hline
	& & \multicolumn{2}{|c|}{$1\sigma$ error} \\
	\cline{3-4}
	Parameter & Fiducial value & $P_g$ [No selection] & $P_g$ [With selection] \\
	\hline
	$b_1$ & $1.5$ & $0.0122$ & $0.0332$ \\
	$b_{\lapl\delta} \[ h^{-2} \, {\rm Mpc}^2 \]$ & $-1$ & $1.78$ & $6.62$ \\
	$\bI \[ h^{-2} \, {\rm Mpc}^2 \]$ & $1$ & $1.53$ & $19.1$ \\
	$P_{\epsilon}^{\{0\}} \[ h^{-3} \, {\rm Mpc}^3 \]$ & $1/n_g$ & $0.131 \, n_g^{-1}$ & $0.330 \, n_g^{-1}$ \\
	$P_{\epsilon}^{\{2\}} \[ h^{-5} \, {\rm Mpc}^5 \]$ & $0$ & $8.83 \times 10^{-4} \, n_g^{-5/3}$ & $2.22 \times 10^{-3} \, n_g^{-5/3}$ \\
	$P_{\eps\varepsilon_{\eta}}^{\{2\}} \[ h^{-5} \, {\rm Mpc}^5 \]$ & $0$ & $5.45 \times 10^{-4} \, n_g^{-5/3}$ & $4.25 \times 10^{-3} \, n_g^{-5/3}$ \\
	$b_2$ & $-0.69$ & $0.577$ & $6.19$ \\
	$b_{K^2}$ & $-0.14$ & $0.651$ & $20.6$ \\
	$b_{\rm td}$ & $(23/42)(b_1-1)$ & $1.41$ & $51.9$ \\
	$b_{\eta}$ & $-1$ & -- & $0.950$ \\
	$\bII \[ h^{-2} \, {\rm Mpc}^2 \]$ & $0$ & -- & $5.56$ \\
	$b_{\Pi_{\parallel}^{[2]}}$ & $0$ & -- & $2.88$ \\
	$b_{(KK)_{\parallel}}$ & $0$ & -- & $39.1$ \\
	$b_{\delta\eta}$ & $-b_1$ & -- & $10.3$ \\
	$b_{\eta^2}$ & $1$ & -- & $5.53$ \\
	$b_{\delta \Pi_{\parallel}^{[2]}}$ & $0$ & -- & $7.60$ \\
	$b_{\eta \Pi_{\parallel}^{[2]}}$ & $0$ & -- & $39.5$ \\
	$b_{(\Pi^{[2]} K)_{\parallel}}$ & $0$ & -- & $86.7$ \\
	$b_{\Pi_{\parallel}^{[3]}}$ & $0$ & -- & -- \\
	\hline
	$\ln D_A \, [\%]$ & $D_A = 1.21 \times 10^3 \, h^{-1} \, {\rm Mpc}$ & $0.197$ & $0.266$ \\
	$\ln H \, [\%]$ & $H = 222 \, h \, {\rm km \, s^{-1} \, Mpc^{-1}}$ & $0.252$ & $0.309$ \\
	$f$ & 0.921 & $0.0108$ & $0.883$ \\
	\hline
	\end{tabular}
\caption{Fiducial values and projected uncertainties for a Euclid-like survey using the power spectrum alone when selection effects are ignored, that is all selection bias parameters are set to the fiducial values given here but not marginalized over, and when selection effects are included.}
\label{table:euclid1}
\end{center}
\end{table}

\afterpage{\clearpage}

\begin{table}[!h]
  \centering
\begin{center}
	\begin{tabular}{|c|c|c|c|}
	\hline
	& & \multicolumn{2}{|c|}{$1\sigma$ error} \\
	\cline{3-4}
	Parameter & Fiducial value & $P_g$ & $P_g + B_g$ \\
	\hline
	$b_1$ & $1.5$ & $0.0213$ & $9.29 \times 10^{-3}$ \\
	$b_{\lapl\delta} \[ h^{-2} \, {\rm Mpc}^2 \]$ & $-1$ & $2.15$ & $0.809$ \\
	$\bI \[ h^{-2} \, {\rm Mpc}^2 \]$ & $1$ & $2.09$ & $0.609$ \\
	$P_{\epsilon}^{\{0\}} \[ h^{-3} \, {\rm Mpc}^3 \]$ & $1/n_g$ & $0.289 \, n_g^{-1}$ & $0.0359 \, n_g^{-1}$ \\
	$P_{\epsilon}^{\{2\}} \[ h^{-5} \, {\rm Mpc}^5 \]$ & $0$ & $9.97 \times 10^{-4} \, n_g^{-5/3}$ & $4.19 \times 10^{-4} \, n_g^{-5/3}$ \\
	$P_{\eps\varepsilon_{\eta}}^{\{2\}} \[ h^{-5} \, {\rm Mpc}^5 \]$ & $0$ & $6.00 \times 10^{-4} \, n_g^{-5/3}$ & $5.19 \times 10^{-4} \, n_g^{-5/3}$ \\
	$b_2$ & $-0.69$ & $0.769$ & $0.0613$ \\
	$b_{K^2}$ & $-0.14$ & $1.19$ & $0.0244$ \\
	$b_{\rm td}$ & $(23/42)(b_1-1)$ & $2.57$ & $0.191$ \\
	$B_{\epsilon}^{\{0\}} \[ h^{-6} \, {\rm Mpc}^6 \]$ & $(1/n_g)^2$ & -- & $0.881 \, n_g^{-2}$ \\
	$P_{\epsilon\epsilon_{\delta}}^{\{0\}} \[ h^{-3} \, {\rm Mpc}^3 \]$ & $2b_1/n_g$ & -- & $0.138 \, n_g^{-1}$ \\
	$P_{\epsilon\epsilon_{\eta}}^{\{0\}} \[ h^{-3} \, {\rm Mpc}^3 \]$ & $0$ & -- & $0.0667 \, n_g^{-1}$ \\
	\hline
	$\ln D_A \, [\%]$ & $D_A = 1.21 \times 10^3 \, h^{-1} \, {\rm Mpc}$ & $0.860$ & $0.342$ \\
	$\ln H \, [\%]$ & $H = 222 \, h \, {\rm km \, s^{-1} \, Mpc^{-1}}$ & $0.818$ & $0.295$ \\
	$\Omega_{c0}$ & 0.258 & $0.0101$ & $3.26 \times 10^{-3}$ \\
	$f$ & 0.921 & $0.0257$ & $6.60 \times 10^{-3}$ \\
	$n_s$ & 0.967 & $0.0288$ & $0.0121$ \\
	$n_{\rm run}$ & 0 & $0.0167$ & $8.28 \times 10^{-3}$ \\
	\hline
	\end{tabular}
\caption{Fiducial values and projected uncertainties for a Euclid-like survey ignoring selection effects, that is setting all selection bias parameters to the fiducial values given in table\ \ref{table:euclid3} but not marginalizing over them.}
\label{table:euclid2}
\end{center}
\end{table}

\afterpage{\clearpage}

\begin{table}[!h]
  \centering
\begin{center}
	\begin{tabular}{|c|c|c|c|}
	\hline
	& & \multicolumn{2}{|c|}{$1\sigma$ error} \\
	\cline{3-4}
	Parameter & Fiducial value & $P_g$ & $P_g + B_g$ \\
	\hline
	$b_1$ & $1.5$ & $0.146$ & $0.0113$ \\
	$b_{\lapl\delta} \[ h^{-2} \, {\rm Mpc}^2 \]$ & $-1$ & $8.86$ & $1.81$ \\
	$\bI \[ h^{-2} \, {\rm Mpc}^2 \]$ & $1$ & $25.1$ & $7.65$ \\
	$P_{\epsilon}^{\{0\}} \[ h^{-3} \, {\rm Mpc}^3 \]$ & $1/n_g$ & $1.10 \, n_g^{-1}$ & $0.0540 \, n_g^{-1}$ \\
	$P_{\epsilon}^{\{2\}} \[ h^{-5} \, {\rm Mpc}^5 \]$ & $0$ & $2.74 \times 10^{-3} \, n_g^{-5/3}$ & $1.05 \times 10^{-3} \, n_g^{-5/3}$ \\
	$P_{\eps\varepsilon_{\eta}}^{\{2\}} \[ h^{-5} \, {\rm Mpc}^5 \]$ & $0$ & $4.89 \times 10^{-3} \, n_g^{-5/3}$ & $3.01 \times 10^{-3} \, n_g^{-5/3}$ \\
	$b_2$ & $-0.69$ & $7.62$ & $0.0925$ \\
	$b_{K^2}$ & $-0.14$ & $30.5$ & $0.0278$ \\
	$b_{\rm td}$ & $(23/42)(b_1-1)$ & $67.6$ & $7.35$ \\
	$b_{\eta}$ & $-1$ & $1.16$ & $0.108$ \\
	$\bII \[ h^{-2} \, {\rm Mpc}^2 \]$ & $0$ & $5.98$ & $3.13$ \\
	$b_{\Pi_{\parallel}^{[2]}}$ & $0$ & $3.65$ & $0.0744$ \\
	$b_{(KK)_{\parallel}}$ & $0$ & $52.8$ & $0.148$ \\
	$b_{\delta\eta}$ & $-b_1$ & $11.7$ & $0.175$ \\
	$b_{\eta^2}$ & $1$ & $5.91$ & $0.190$ \\
	$b_{\delta \Pi_{\parallel}^{[2]}}$ & $0$ & $8.78$ & $3.93$ \\
	$b_{\eta \Pi_{\parallel}^{[2]}}$ & $0$ & $51.4$ & $15.5$ \\
	$b_{(\Pi^{[2]} K)_{\parallel}}$ & $0$ & $99.6$ & $15.8$ \\
	$b_{\Pi_{\parallel}^{[3]}}$ & $0$ & -- & -- \\
	$B_{\epsilon}^{\{0\}} \[ h^{-6} \, {\rm Mpc}^6 \]$ & $(1/n_g)^2$ & -- & $1.04 \, n_g^{-2}$ \\
	$P_{\epsilon\epsilon_{\delta}}^{\{0\}} \[ h^{-3} \, {\rm Mpc}^3 \]$ & $2b_1/n_g$ & -- & $0.167 \, n_g^{-1}$ \\
	$P_{\epsilon\epsilon_{\eta}}^{\{0\}} \[ h^{-3} \, {\rm Mpc}^3 \]$ & $0$ & -- & $0.224 \, n_g^{-1}$ \\
	\hline
	$\ln D_A \, [\%]$ & $D_A = 1.21 \times 10^3 \, h^{-1} \, {\rm Mpc}$ & $2.17$ & $0.359$ \\
	$\ln H \, [\%]$ & $H = 222 \, h \, {\rm km \, s^{-1} \, Mpc^{-1}}$ & $1.81$ & $0.379$ \\
	$\Omega_{c0}$ & $0.258$ & $0.0235$ & $3.50 \times 10^{-3}$ \\
	$f$ & 0.921 & $1.08$ & $0.0899$ \\
	$n_s$ & 0.967 & $0.0874$ & $0.0177$ \\
	$n_{\rm run}$ & 0 & $0.0791$ & $0.0115$ \\
	\hline
	\end{tabular}
        \caption{Fiducial values and projected uncertainties for a Euclid-like survey including selection effects. Among the cosmological parameters, the constraints on $f$ degrade most, by about a factor of 14 in the case of the combined power spectrum and bispectrum analysis. For the power spectrum only, the inclusion of selection effects removes all constraining power on $f$.}
\label{table:euclid3}
\end{center}
\end{table}

\afterpage{\clearpage}

\begin{table}[!h]
  \centering
\begin{center}
\begin{tabular}{|c|c|c|c|c|}
\hline
& \multicolumn{2}{|c|}{Shift, $|\Delta\theta^a|/\sigma^a$} & \multicolumn{2}{|c|}{Shift, $|\Delta\theta^a|/\sigma^a$}  \\
Parameter & \multicolumn{2}{|c|}{$\Delta b_{\eta}/b_{\eta} = 1\%$} & \multicolumn{2}{|c|}{$\Delta b_{\rm td}/b_{\rm td} = 10\%$} \\
\cline{2-5}
& $P_g$ & $P_g + B_g$ & $P_g$ & $P_g + B_g$ \\
\hline
$b_1$ & $0.0172$ & $0.117$ & $0.0182$ & $0.227$ \\
$b_{\lapl\delta} \[ h^{-2} \, {\rm Mpc}^2 \]$ & $0.0840$ & $0.107$ & $8.05 \times 10^{-3}$ & $0.142$ \\
$\bI \[ h^{-2} \, {\rm Mpc}^2 \]$ & $0.130$ & $0.0474$ & $0.0164$ & $0.0880$ \\
$P_{\epsilon}^{\{0\}} \[ h^{-3} \, {\rm Mpc}^3 \]$ & $0.0684$ & $0.0660$ & $0.0661$ & $0.0779$ \\
$P_{\epsilon}^{\{2\}} \[ h^{-1} \, {\rm Mpc} \]$ & $0.0537$ & $0.177$ & $2.31 \times 10^{-3}$ & $0.114$ \\
$P_{\eps\varepsilon_{\eta}}^{\{2\}} \[ h^{-1} \, {\rm Mpc} \]$ & $0.0157$ & $0.0522$ & $1.10 \times 10^{-3}$ & $0.0112$ \\
$b_2$ & $0.156$ & $0.203$ & $0.0160$ & $0.0505$ \\
$b_{K^2}$ & $0.0974$ & $0.0146$ & $0.117$ & $0.0849$ \\
$b_{\rm td}$ & $0.0864$ & $0.103$ & -- & -- \\
$B_{\epsilon}^{\{0\}} \[ h^{-6} \, {\rm Mpc}^6 \]$ & -- & $0.0704$ & -- & $5.88 \times 10^{-3}$ \\
$P_{\epsilon\epsilon_{\delta}}^{\{0\}} \[ h^{-3} \, {\rm Mpc}^3 \]$ & -- & $0.156$ & -- & $0.0479$ \\
$P_{\epsilon\epsilon_{\eta}}^{\{0\}} \[ h^{-3} \, {\rm Mpc}^3 \]$ & -- & $0.217$ & -- & $0.0438$ \\
\hline
$\ln D_A$ & $0.0336$ & $0.0440$ & $0.0138$ & $0.0133$ \\
$\ln H$ & $0.0503$ & $0.0734$ & $0.0128$ & $0.0319$ \\
$\Omega_{c0}$ & $0.0466$ & $0.0445$ & $0.0152$ & $0.0307$ \\
$f$ & $0.270$ & $1.23$ & $0.0266$ & $0.0191$ \\
$n_s$ & $0.137$ & $0.130$ & $0.0123$ & $0.0954$ \\
$n_{\rm run}$ & $0.107$ & $0.168$ & $9.45 \times 10^{-4}$ & $0.101$ \\
\hline
\end{tabular}
\caption{(Columns 2\,--\,3) Parameter shifts, compared to the corresponding $1\sigma$ errors, for a Euclid-like survey when selection effects are ignored and $b_{\eta}$ differs from its true value by $1\%$. (Columns 4\,--\,5) Parameter shifts when selection effects and $b_{\rm td}$ are ignored and $b_{\rm td}$ differs from its true value by $10\%$.}
\label{table:euclid4}
\end{center}
\end{table}


\newpage

\bibliography{references}

\providecommand{\href}[2]{#2}\begingroup\raggedright\begin{thebibliography}{10}

\bibitem{Bennett:2012zja}
C.~L. Bennett et~al., {\it {Nine-year Wilkinson Microwave Anisotropy Probe
  (WMAP) observations: Final maps and results}},  {\em Astrophys. J. Suppl.}
  {\bf 208} (2013) 20, [\href{http://arxiv.org/abs/1212.5225}{{\tt
  arXiv:1212.5225}}].

\bibitem{Alam:2016hwk}
S.~Alam et~al., {\it {The clustering of galaxies in the completed SDSS-III
  Baryon Oscillation Spectroscopic Survey: Cosmological analysis of the DR12
  galaxy sample}},  {\em Mon. Not. Roy. Astron. Soc.} {\bf 470} (2017)
  2617--2652, [\href{http://arxiv.org/abs/1607.03155}{{\tt arXiv:1607.03155}}].

\bibitem{Abbott:2017wau}
T.~M.~C. Abbott et~al., {\it {Dark Energy Survey year 1 results: Cosmological
  constraints from galaxy clustering and weak lensing}},  {\em Phys. Rev.} {\bf
  D98} (2018) 043526, [\href{http://arxiv.org/abs/1708.01530}{{\tt
  arXiv:1708.01530}}].

\bibitem{Aghanim:2018eyx}
N.~Aghanim et~al., {\it {Planck 2018 results. VI. Cosmological parameters}},
  \href{http://arxiv.org/abs/1807.06209}{{\tt arXiv:1807.06209}}.

\bibitem{Bernardeau:2001qr}
F.~Bernardeau, S.~Colombi, E.~Gazta\~{n}aga, and R.~Scoccimarro, {\it {Large
  scale structure of the Universe and cosmological perturbation theory}},  {\em
  Phys. Rept.} {\bf 367} (2002) 1--248,
  [\href{http://arxiv.org/abs/astro-ph/0112551}{{\tt astro-ph/0112551}}].

\bibitem{Desjacques:2016bnm}
V.~Desjacques, D.~Jeong, and F.~Schmidt, {\it {Large-scale galaxy bias}},  {\em
  Phys. Rept.} {\bf 733} (2018) 1--193,
  [\href{http://arxiv.org/abs/1611.09787}{{\tt arXiv:1611.09787}}].

\bibitem{Zheng:2010jf}
Z.~Zheng, R.~Cen, H.~Trac, and J.~Miralda-Escude, {\it {Radiative transfer
  modeling of Ly{$\alpha$} emitters. II. New effects in galaxy clustering}},
  {\em Astrophys. J.} {\bf 726} (2010) 38--64,
  [\href{http://arxiv.org/abs/1003.4990}{{\tt arXiv:1003.4990}}].

\bibitem{Wyithe:2011mt}
S.~Wyithe and M.~Dijkstra, {\it {Non-gravitational contributions to the
  clustering of Ly{$\alpha$} selected galaxies: Implications for cosmological
  surveys}},  {\em Mon. Not. Roy. Astron. Soc.} {\bf 415} (2011) 3929,
  [\href{http://arxiv.org/abs/1104.0712}{{\tt arXiv:1104.0712}}].

\bibitem{Behrens:2017xmm}
C.~Behrens, C.~Byrohl, S.~Saito, and J.~Niemeyer, {\it {The impact of
  Lyman-$\alpha$ radiative transfer on large-scale clustering in the Illustris
  simulation}},  {\em Astron. Astrophys.} {\bf 614} (2018) A31,
  [\href{http://arxiv.org/abs/1710.06171}{{\tt arXiv:1710.06171}}].

\bibitem{Hirata:2009qz}
C.~M. Hirata, {\it {Tidal alignments as a contaminant of redshift space
  distortions}},  {\em Mon. Not. Roy. Astron. Soc.} {\bf 399} (2009) 1074,
  [\href{http://arxiv.org/abs/0903.4929}{{\tt arXiv:0903.4929}}].

\bibitem{Krause:2010tt}
E.~Krause and C.~Hirata, {\it {Tidal alignments as a contaminant of the galaxy
  bispectrum}},  {\em Mon. Not. Roy. Astron. Soc.} {\bf 410} (2011) 2730,
  [\href{http://arxiv.org/abs/1004.3611}{{\tt arXiv:1004.3611}}].

\bibitem{Martens:2018uqj}
D.~Martens, C.~M. Hirata, A.~J. Ross, and X.~Fang, {\it {A radial measurement
  of the galaxy tidal alignment magnitude with BOSS data}},  {\em Mon. Not.
  Roy. Astron. Soc.} {\bf 478} (2018) 711--732,
  [\href{http://arxiv.org/abs/1802.07708}{{\tt arXiv:1802.07708}}].

\bibitem{Obuljen:2019ukz}
A.~Obuljen, N.~Dalal, and W.~J. Percival, {\it {Anisotropic halo assembly bias
  and redshift-space distortions}},  {\em JCAP} {\bf 1910} (2019) 020,
  [\href{http://arxiv.org/abs/1906.11823}{{\tt arXiv:1906.11823}}].

\bibitem{Desjacques:2018pfv}
V.~Desjacques, D.~Jeong, and F.~Schmidt, {\it {The galaxy power spectrum and
  bispectrum in redshift space}},  {\em JCAP} {\bf 1812} (2018) 035,
  [\href{http://arxiv.org/abs/1806.04015}{{\tt arXiv:1806.04015}}].

\bibitem{Porto:2016pyg}
R.~A. Porto, {\it {The effective field theorist's approach to gravitational
  dynamics}},  {\em Phys. Rept.} {\bf 633} (2016) 1--104,
  [\href{http://arxiv.org/abs/1601.04914}{{\tt arXiv:1601.04914}}].

\bibitem{Baumann:2010tm}
D.~Baumann, A.~Nicolis, L.~Senatore, and M.~Zaldarriaga, {\it {Cosmological
  non-linearities as an effective fluid}},  {\em JCAP} {\bf 1207} (2012) 051,
  [\href{http://arxiv.org/abs/1004.2488}{{\tt arXiv:1004.2488}}].

\bibitem{Carrasco:2012cv}
J.~J.~M. Carrasco, M.~P. Hertzberg, and L.~Senatore, {\it {The effective field
  theory of cosmological large scale structures}},  {\em JHEP} {\bf 1209}
  (2012) 082, [\href{http://arxiv.org/abs/1206.2926}{{\tt arXiv:1206.2926}}].

\bibitem{Senatore:2014eva}
L.~Senatore, {\it {Bias in the effective field theory of large scale
  structures}},  {\em JCAP} {\bf 1511} (2015) 007,
  [\href{http://arxiv.org/abs/1406.7843}{{\tt arXiv:1406.7843}}].

\bibitem{Senatore:2014vja}
L.~Senatore and M.~Zaldarriaga, {\it {Redshift space distortions in the
  effective field theory of large scale structures}},
  \href{http://arxiv.org/abs/1409.1225}{{\tt arXiv:1409.1225}}.

\bibitem{Mirbabayi:2014zca}
M.~Mirbabayi, F.~Schmidt, and M.~Zaldarriaga, {\it {Biased tracers and time
  evolution}},  {\em JCAP} {\bf 1507} (2015) 030,
  [\href{http://arxiv.org/abs/1412.5169}{{\tt arXiv:1412.5169}}].

\bibitem{Kaiser:1987qv}
N.~Kaiser, {\it {Clustering in real space and in redshift space}},  {\em Mon.
  Not. Roy. Astron. Soc.} {\bf 227} (1987) 1--27.

\bibitem{Euclid}
R.~Laureijs et~al., {\it {Euclid definition study report}},
  \href{http://arxiv.org/abs/1110.3193}{{\tt arXiv:1110.3193}}.

\bibitem{Green:2012mj}
J.~Green et~al., {\it {Wide-Field InfraRed Survey Telescope (WFIRST) final
  report}},  \href{http://arxiv.org/abs/1208.4012}{{\tt arXiv:1208.4012}}.

\bibitem{Hill:2008mv}
G.~J. Hill, K.~Gebhardt, E.~Komatsu, N.~Drory, P.~J. MacQueen, et~al., {\it
  {The Hobby-Eberly Telescope Dark Energy Experiment (HETDEX): Description and
  early pilot survey results}},  {\em ASP Conf. Ser.} {\bf 399} (2008)
  115--118, [\href{http://arxiv.org/abs/0806.0183}{{\tt arXiv:0806.0183}}].

\bibitem{V2}
S.~Dodelson and F.~Schmidt, {\em {Modern Cosmology, Second Edition}}.
\newblock Academic Press, San Diego, CA, 2020.

\bibitem{Goroff:1986ep}
M.~H. Goroff, B.~Grinstein, S.-J. Rey, and M.~B. Wise, {\it {Coupling of modes
  of cosmological mass density fluctuations}},  {\em Astrophys. J.} {\bf 311}
  (1986) 6--14.

\bibitem{Heavens:1998es}
A.~F. Heavens, S.~Matarrese, and L.~Verde, {\it {The non-linear redshift space
  power spectrum of galaxies}},  {\em Mon. Not. Roy. Astron. Soc.} {\bf 301}
  (1998) 797--808, [\href{http://arxiv.org/abs/astro-ph/9808016}{{\tt
  astro-ph/9808016}}].

\bibitem{Takahashi:2008yk}
R.~Takahashi, {\it {Third order density perturbation and one-loop power
  spectrum in a dark energy dominated Universe}},  {\em Prog. Theor. Phys.}
  {\bf 120} (2008) 549--559, [\href{http://arxiv.org/abs/0806.1437}{{\tt
  arXiv:0806.1437}}].

\bibitem{Carrasco:2013mua}
J.~J.~M. Carrasco, S.~Foreman, D.~Green, and L.~Senatore, {\it {The effective
  field theory of large scale structures at two loops}},  {\em JCAP} {\bf 1407}
  (2014) 057, [\href{http://arxiv.org/abs/1310.0464}{{\tt arXiv:1310.0464}}].

\bibitem{Paech:2016hod}
K.~Paech, N.~Hamaus, B.~Hoyle, M.~Costanzi, T.~Giannantonio, S.~Hagstotz,
  G.~Sauerwein, and J.~Weller, {\it {Cross-correlation of galaxies and galaxy
  clusters in the Sloan Digital Sky Survey and the importance of non-Poissonian
  shot noise}},  {\em Mon. Not. Roy. Astron. Soc.} {\bf 470} (2017) 2566--2577,
  [\href{http://arxiv.org/abs/1612.02018}{{\tt arXiv:1612.02018}}].

\bibitem{Hamaus:2010im}
N.~Hamaus, U.~Seljak, V.~Desjacques, R.~E. Smith, and T.~Baldauf, {\it
  {Minimizing the stochasticity of halos in large-scale structure surveys}},
  {\em Phys. Rev. D} {\bf 82} (2010) 043515,
  [\href{http://arxiv.org/abs/1004.5377}{{\tt arXiv:1004.5377}}].

\bibitem{Ginzburg:2017mgf}
D.~Ginzburg, V.~Desjacques, and K.~C. Chan, {\it {Shot noise and biased
  tracers: A new look at the halo model}},  {\em Phys. Rev. D} {\bf 96} (2017)
  083528, [\href{http://arxiv.org/abs/1706.08738}{{\tt arXiv:1706.08738}}].

\bibitem{Desjacques:2009kt}
V.~Desjacques and R.~K. Sheth, {\it {Redshift space correlations and
  scale-dependent stochastic biasing of density peaks}},  {\em Phys. Rev.} {\bf
  D81} (2010) 023526, [\href{http://arxiv.org/abs/0909.4544}{{\tt
  arXiv:0909.4544}}].

\bibitem{Scoccimarro:2004tg}
R.~Scoccimarro, {\it {Redshift-space distortions, pairwise velocities and
  nonlinearities}},  {\em Phys. Rev.} {\bf D70} (2004) 083007,
  [\href{http://arxiv.org/abs/astro-ph/0407214}{{\tt astro-ph/0407214}}].

\bibitem{Seljak:2011tx}
U.~Seljak and P.~McDonald, {\it {Distribution function approach to redshift
  space distortions}},  {\em JCAP} {\bf 1111} (2011) 039,
  [\href{http://arxiv.org/abs/1109.1888}{{\tt arXiv:1109.1888}}].

\bibitem{Perko:2016puo}
A.~Perko, L.~Senatore, E.~Jennings, and R.~H. Wechsler, {\it {Biased tracers in
  redshift space in the EFT of large-scale structure}},
  \href{http://arxiv.org/abs/1610.09321}{{\tt arXiv:1610.09321}}.

\bibitem{Schmidt:2018bkr}
F.~Schmidt, F.~Elsner, J.~Jasche, N.~M. Nguyen, and G.~Lavaux, {\it {A rigorous
  EFT-based forward model for large-scale structure}},  {\em JCAP} {\bf 01}
  (2019) 042, [\href{http://arxiv.org/abs/1808.02002}{{\tt arXiv:1808.02002}}].

\bibitem{Tegmark:1997rp}
M.~Tegmark, {\it {Measuring cosmological parameters with galaxy surveys}},
  {\em Phys. Rev. Lett.} {\bf 79} (1997) 3806--3809,
  [\href{http://arxiv.org/abs/astro-ph/9706198}{{\tt astro-ph/9706198}}].

\bibitem{Dodelson:2005ir}
S.~Dodelson, C.~Shapiro, and M.~J. White, {\it {Reduced shear power spectrum}},
   {\em Phys. Rev.} {\bf D73} (2006) 023009,
  [\href{http://arxiv.org/abs/astro-ph/0508296}{{\tt astro-ph/0508296}}].

\bibitem{Heavens:2009nx}
A.~Heavens, {\it {Statistical techniques in cosmology}},
  \href{http://arxiv.org/abs/0906.0664}{{\tt arXiv:0906.0664}}.

\bibitem{Gebhardt:2018zuj}
H.~S. Grasshorn~Gebhardt et~al., {\it {Unbiased cosmological parameter
  estimation from emission line surveys with interlopers}},  {\em Astrophys.
  J.} {\bf 876} (2019) 32, [\href{http://arxiv.org/abs/1811.06982}{{\tt
  arXiv:1811.06982}}].

\bibitem{Markovic:2019sva}
K.~Markovic, B.~Bose, and A.~Pourtsidou, {\it {Assessing non-linear models for
  galaxy clustering I: Unbiased growth forecasts from multipole expansion}},
  {\em Open J. Astrophys.} {\bf 2} (2019) 13,
  [\href{http://arxiv.org/abs/1904.11448}{{\tt arXiv:1904.11448}}].

\bibitem{GalaxyFormationBook}
H.~{Mo}, F.~C. {van den Bosch}, and S.~{White}, {\em {Galaxy Formation and
  Evolution, First Edition}}.
\newblock Cambridge University Press, 2010.

\bibitem{Behroozi:2012iw}
P.~S. Behroozi, R.~H. Wechsler, and C.~Conroy, {\it {The average star formation
  histories of galaxies in dark matter halos from $z=$0-8}},  {\em Astrophys.
  J.} {\bf 770} (2013) 57, [\href{http://arxiv.org/abs/1207.6105}{{\tt
  arXiv:1207.6105}}].

\bibitem{Pozzetti:2016cch}
L.~Pozzetti, C.~Hirata, J.~Geach, A.~Cimatti, C.~Baugh, O.~Cucciati, A.~Merson,
  P.~Norberg, and D.~Shi, {\it {Modelling the number density of H$\alpha$
  emitters for future spectroscopic near-IR space missions}},  {\em Astron.
  Astrophys.} {\bf 590} (2016) A3, [\href{http://arxiv.org/abs/1603.01453}{{\tt
  arXiv:1603.01453}}].

\bibitem{Merson:2017efv}
A.~Merson, Y.~Wang, A.~Benson, A.~Faisst, D.~Masters, A.~Kiessling, and
  J.~Rhodes, {\it {Predicting H$\alpha$ emission-line galaxy counts for future
  galaxy redshift surveys}},  {\em Mon. Not. Roy. Astron. Soc.} {\bf 474}
  (2018) 177--196, [\href{http://arxiv.org/abs/1710.00833}{{\tt
  arXiv:1710.00833}}].

\bibitem{khostovan/etal:2019}
A.~A. Khostovan et~al., {\it {The clustering of typical Ly{\ensuremath{\alpha}}
  emitters from z {\ensuremath{\sim}} 2.5-6: Host halo masses depend on
  Ly{\ensuremath{\alpha}} and UV luminosities}},  {\em \mnras} {\bf 489} (2019)
  555--573, [\href{http://arxiv.org/abs/1811.00556}{{\tt arXiv:1811.00556}}].

\bibitem{Adam:2015rua}
R.~Adam et~al., {\it {Planck 2015 results. I. Overview of products and
  scientific results}},  \href{http://arxiv.org/abs/1502.01582}{{\tt
  arXiv:1502.01582}}.

\bibitem{Ade:2015xua}
P.~A.~R. Ade et~al., {\it {Planck 2015 results. XIII. Cosmological
  parameters}},  \href{http://arxiv.org/abs/1502.01589}{{\tt
  arXiv:1502.01589}}.

\bibitem{Scoccimarro:1999ed}
R.~Scoccimarro, H.~Couchman, and J.~A. Frieman, {\it {The bispectrum as a
  signature of gravitational instability in redshift-space}},  {\em Astrophys.
  J.} {\bf 517} (1999) 531--540,
  [\href{http://arxiv.org/abs/astro-ph/9808305}{{\tt astro-ph/9808305}}].

\bibitem{Gil-Marin:2014pva}
H.~Gil-Mar{\'\i}n, C.~Wagner, J.~Nore{\~n}a, L.~Verde, and W.~Percival, {\it
  {Dark matter and halo bispectrum in redshift space: Theory and
  applications}},  {\em JCAP} {\bf 12} (2014) 029,
  [\href{http://arxiv.org/abs/1407.1836}{{\tt arXiv:1407.1836}}].

\bibitem{Sanchez:2020vvb}
A.~G. Sanchez, {\it {Let us bury the prehistoric $h$: Arguments against using
  $h^{-1} \, Mpc$ units in observational cosmology}},
  \href{http://arxiv.org/abs/2002.07829}{{\tt arXiv:2002.07829}}.

\bibitem{Sefusatti:2006pa}
E.~Sefusatti, M.~Crocce, S.~Pueblas, and R.~Scoccimarro, {\it {Cosmology and
  the bispectrum}},  {\em Phys. Rev. D} {\bf 74} (2006) 023522,
  [\href{http://arxiv.org/abs/astro-ph/0604505}{{\tt astro-ph/0604505}}].

\bibitem{Yankelevich:2018uaz}
V.~Yankelevich and C.~Porciani, {\it {Cosmological information in the
  redshift-space bispectrum}},  {\em Mon. Not. Roy. Astron. Soc.} {\bf 483}
  (2019) 2078--2099, [\href{http://arxiv.org/abs/1807.07076}{{\tt
  arXiv:1807.07076}}].

\bibitem{Gualdi:2020ymf}
D.~Gualdi and L.~Verde, {\it {Galaxy redshift-space bispectrum: The importance
  of being anisotropic}},  {\em JCAP} {\bf 06} (2020) 041,
  [\href{http://arxiv.org/abs/2003.12075}{{\tt arXiv:2003.12075}}].

\bibitem{Scoccimarro:2000sn}
R.~Scoccimarro, {\it {The bispectrum: From theory to observations}},  {\em
  Astrophys. J.} {\bf 544} (2000) 597,
  [\href{http://arxiv.org/abs/astro-ph/0004086}{{\tt astro-ph/0004086}}].

\bibitem{Wadekar:2019rdu}
D.~Wadekar and R.~Scoccimarro, {\it {The galaxy power spectrum multipoles
  covariance in perturbation theory}},
  \href{http://arxiv.org/abs/1910.02914}{{\tt arXiv:1910.02914}}.

\bibitem{Blot:2018oxk}
L.~Blot et~al., {\it {Comparing approximate methods for mock catalogues and
  covariance matrices II: Power spectrum multipoles}},  {\em Mon. Not. Roy.
  Astron. Soc.} {\bf 485} (2019) 2806--2824,
  [\href{http://arxiv.org/abs/1806.09497}{{\tt arXiv:1806.09497}}].

\bibitem{Vogeley:1996xu}
M.~S. Vogeley and A.~S. Szalay, {\it {Eigenmode analysis of galaxy redshift
  surveys I. Theory and methods}},  {\em Astrophys. J.} {\bf 465} (1996)
  34--53, [\href{http://arxiv.org/abs/astro-ph/9601185}{{\tt
  astro-ph/9601185}}].

\bibitem{Byun:2017fkz}
J.~Byun, A.~Eggemeier, D.~Regan, D.~Seery, and R.~E. Smith, {\it {Towards
  optimal cosmological parameter recovery from compressed bispectrum
  statistics}},  {\em Mon. Not. Roy. Astron. Soc.} {\bf 471} (2017) 1581--1618,
  [\href{http://arxiv.org/abs/1705.04392}{{\tt arXiv:1705.04392}}].

\bibitem{Gualdi:2017iey}
D.~Gualdi, M.~Manera, B.~Joachimi, and O.~Lahav, {\it {Maximal compression of
  the redshift space galaxy power spectrum and bispectrum}},  {\em Mon. Not.
  Roy. Astron. Soc.} {\bf 476} (2018) 4045--4070,
  [\href{http://arxiv.org/abs/1709.03600}{{\tt arXiv:1709.03600}}].

\bibitem{Shoji:2008xn}
M.~Shoji, D.~Jeong, and E.~Komatsu, {\it {Extracting angular diameter distance
  and expansion rate of the Universe from two-dimensional galaxy power spectrum
  at high redshifts: Baryon acoustic oscillation fitting versus full
  modeling}},  {\em Astrophys. J.} {\bf 693} (2009) 1404--1416,
  [\href{http://arxiv.org/abs/0805.4238}{{\tt arXiv:0805.4238}}].

\bibitem{Greig:2012zw}
B.~Greig, E.~Komatsu, and J.~S.~B. Wyithe, {\it {Cosmology from clustering of
  Ly$\alpha$ galaxies: Breaking non-gravitational Ly$\alpha$ radiative transfer
  degeneracies using the bispectrum}},  {\em Mon. Not. Roy. Astron. Soc.} {\bf
  431} (2013) 1777, [\href{http://arxiv.org/abs/1212.0977}{{\tt
  arXiv:1212.0977}}].

\end{thebibliography}\endgroup
\bibliographystyle{JCAP}

\end{document}